\documentclass[num-refs]{wiley-article}

\usepackage{siunitx}
\usepackage{subcaption}

\papertype{Original Article}
\paperfield{Journal Section}

\title{The helix approach: using dynamic individual pitch control to enhance wake mixing in wind farms}

\author[1]{Joeri A. Frederik}
\author[1]{Bart M. Doekemeijer}
\author[1]{Sebastiaan P. Mulders}
\author[1]{Jan-Willem van Wingerden}

\affil[1]{Delft Center of Systems and Control, TU Delft, Delft, 2628 CD, The Netherlands}

\corraddress{J.A. Frederik, Delft Center of Systems and Control, TU Delft, Delft, 2628 CD, The Netherlands}
\corremail{j.a.frederik@tudelft.nl}

\runningauthor{Frederik et al.}

\begin{document}

\maketitle

\begin{abstract}
Wind farm control using dynamic concepts is a research topic that is receiving an increasing amount of interest. The main concept of this approach is that dynamic variations of the wind turbine control settings lead to higher wake turbulence, and subsequently faster wake recovery due to increased mixing. As a result, downstream turbines experience higher wind speeds, thus increasing their energy capture. The current state of the art in dynamic wind farm control is to vary the magnitude of the thrust force of an upstream turbine. Although very effective, this approach also leads to increased power and thrust variations, negatively impacting energy quality and fatigue loading. In this paper, a novel approach for the dynamic control of wind turbines in a wind farm is proposed: using individual pitch control, the fixed-frame tilt and yaw moments on the turbine are varied, thus dynamically manipulating the wake. This strategy is named the \textit{helix approach} since the resulting wake has a helical shape. Large eddy simulations of a two-turbine wind farm show that the helix approach leads to enhanced wake mixing with minimal power and thrust variations. 


\end{abstract}

\section{Introduction}
The interaction between wind turbines in a wind farm through their wakes is a phenomenon that has been studied for decades \citep{windfarm1979,jensen,katic1987}, and is still a relevant topic today \citep{bastankhah,annonifloris}. For the purpose of power maimization and load minimization, this interaction can be manipulated using techniques from the control engineering community. A comprehensive survey on wind farm modelling and control can be found in \cite{boersma2017}. In wind farm control, two different approaches can be distinguished: induction control (sometimes called derating control) and wake redirection control (sometimes called wake steering). The former approach uses the induction factor, i.e., the in-wake velocity deficit, of the turbines as control input, whereas the latter approach exploits the yaw angle of turbines. Both approaches follow the same strategy: the upstream machines in a wind turbine array will lose power due to locally suboptimal induction or yaw settings, and downstream machines experience higher wind speeds which increases their power production. 

The examples of induction control and wake redirection control are plentiful. Induction control has shown promising results in different simulation environments using model-free optimization \citep{marden,rotea} or Model Predictive Control \citep{vali}. However, recent studies with high-fidelity simulation models \citep{annoni}, scaled wind tunnel experiments \citep{Campagnolo2016a} and full-scale experiments \citep{vdhoek2019} indicate that the potential wind farm power gain of induction control is minor to non-existent. Therefore, the focus in the literature for power maximization in wind farms is shifted towards wake redirection. Wake deflection through yaw is first modelled in \citep{jimenez2010}, and is also investigated on full-scale turbines using lidar measurements \citep{raach2016}. Both scaled wind tunnel experiments \citep{Campagnolo2016_WFC} and full-scale tests \citep{fleming2017,howland} indicate that this strategy can effectively increase the power production of a wind farm. All these references have in common that they focus on steady-state optimal control of a wind farm. Therefore, time-varying control inputs that purposely influence the inherently dynamic nature of the wind are disregarded.

To the best of the authors' knowledge, the first mention of dynamic control being used to increase the performance of wind farms is in an industrial patent \citep{dicpatent}. This patent describes different control methods involving either dynamic induction, dynamic yawing or wake deformation through cyclic pitch signals. What these control methods have in common, is that they aim to increase wake mixing by changing the control inputs over time. Wake mixing is the phenomenon where the wake interacts with the adjacent, higher velocity, free-stream flow. As a result, the wake recovers some of the energy extracted by the upstream turbine, such that a downstream turbine experiences a higher wind velocity. However, only the general idea is described; no experiments or simulations are performed, and the effectiveness of these methods is not evaluated.

Recently, dynamic wind farm control has gained interest in the scientific field. Dynamic Induction Control (DIC) specifically is a research topic that has seen a number of publications studying its potential in simulations \citep{Munters:2016,Munters:2017,Munters:2018} and in scaled wind tunnel experiments \citep{frederik2019}. To enable practical implementation, the most recent papers focus on a smaller subset of dynamic signals, namely sinusoidal signals \citep{Munters:2018}. In \citep{Munters:2018}, a grid search is performed using Large Eddy Simulations (LES) to determine the amplitude and frequency of the sinusoidal excitation that maximize the farm-wide power production. In \citep{frederik2019}, wind tunnel experiments are performed to validate this approach, showing positive results. A different dynamic control approach is investigated using high-fidelity simulations in \citep{dicyaw}. Here, the yaw angle of a turbine is varied sinusoidally, such that increased wake meandering is induced.

The above-mentioned approaches do have an important drawback: because of the varying induction factor or yaw angle signals of the upstream turbine, the thrust force on the rotor varies significantly. As a result, this turbine experiences substantial power and load fluctuations, which is disadvantageous from a power quality perspective. In this paper, a novel approach to dynamic wake mixing is introduced, which is expected to lead to lower power and thrust variations. This approach makes use of Individual Pitch Control (IPC), a procedure in which the blade pitch angles of a wind turbine are controlled independently of each other.

IPC is a well-known strategy in the wind turbine control community. It is usually applied to alleviate periodic loads on turbines with minimal power loss, as first proposed in \citep{bossanyi2003individual,bossanyi2005further}. Further research into load reducing IPC algorithms is still a relevant research direction, for example into using an azimuth offset \citep{mulders2019analysis} or implementing more advanced control strategies \citep{navalkar2014subspace,frederik2018cep,frederik2018torque}. Research where IPC is used to increase the power production of a wind farm is limited. Experiments have been conducted where IPC is used for wake steering \citep{ipcwakesteering} or power maximization in case of partial wake overlap \citep{fleming2015ipc}. However, the results were inconclusive and no further research has been published since.

In this paper, wake steering through individual pitch control is combined with the concept of dynamic wind farm control to forge a novel approach. This approach, called Dynamic IPC (DIPC), uses the Multi-Blade Coordinate (MBC) transformations to vary the tilt and yaw moments on the rotor. Thus, the wake is manipulated, slowly varying its direction over time. This is hypothesized to result in enhanced wake mixing, such that downstream turbines in a wind turbine array can increase their power production with minimal rotor thrust fluctuations. A patent by the authors  describing this concept is pending \citep{ipcpatent}.

The main contributions of this paper are threefold. First of all, the novel DIPC approach is described. Secondly, a specific DIPC strategy called the \textit{helix approach} is defined, which dynamically moves the wake both horizontally and vertically. Finally, the effectiveness of this helix approach is evaluated through high-fidelity simulations. These simulations are executed using the LES code SOWFA \citep{SOWFA_General}. The effects of DIPC both on the wake and on a downstream turbine is investigated. A thorough comparison is made with existing control strategies to evaluate the performance of DIPC.

This paper is organized as follows: in Section~\ref{sec:sowfa}, the simulation environment is defined. Section~\ref{sec:approach} thoroughly describes the working principles of DIPC in general and the helix approach specifically. The potential of this approach as a wind farm control approach will then be demonstrated in Section~\ref{sec:results} through high-fidelity simulations. Finally, conclusions are drawn and future work is discussed in Section~\ref{sec:conclusions}.

\section{Simulation Environment}\label{sec:sowfa}

The proposed control strategy is evaluated in the Simulator fOr Wind Farm Applications (SOWFA) \citep{SOWFA_General}, which is a high-fidelity simulation environment developed by the U.S. National Renewable Energy Laboratory (NREL). SOWFA is a large-eddy solver for the fluid dynamics in the turbulent atmosphere. The interaction with one or multiple wind turbines \cite{Churchfield2012}, accounting for the Coriolis force and Buoyancy effects, is included in SOWFA. Turbines are modelled as actuator disks or actuator lines as demonstrated in \citep{Sorensen2012}. The SOWFA source code was adapted to allow for specifications of a different pitch setpoint for each individual blade, enabling the implementation of Individual Pitch Control (IPC).

In this work, two different simulation cases are defined. First of all, wind with a uniform inflow profile is used to demonstrate the working principles of Dynamic IPC (DIPC). It is recognized that these conditions do not represent realistic working conditions in an actual wind farm. However, due to the absence of turbulence, these simulations are perfectly suited to visualize the effects of DIPC on the wake of a turbine, as presented in Section~\ref{sec:approach}.

\begin{table}[h]
    \centering
    \caption{Numerical simulation scheme in SOWFA for uniform simulations}
    \label{tab:SOWFA}    
    \begin{tabular}{r l l}
          & \textbf{Case I: uniform flow} & \textbf{Case II: turbulent flow} \\ \hline
         Turbine & DTU 10MW \cite{DTU10MW} & DTU 10MW \cite{DTU10MW} \\
         Rotor diameter & $178.3$~m & $178.3$~m\\
         Domain size & $2.5$~km~$\times~1$~km~$\times$~$0.6$~km & $3$~km~$\times~3$~km~$\times$~$1$~km \\
         Cell size (outer region) & $50$~m~$\times~50$~m~$\times$~$50$~m & $10$~m~$\times~10$~m~$\times$~$10$~m \\
         Cell size (near rotor) & $3.125$~m~$\times~3.125$~m~$\times$~$3.125$~m & $1.25$~m~$\times~1.25$~m~$\times$~$1.25$~m\\
         Inflow wind speed & $9.0$~m/s & $9.0$~m/s  \\      
         Inflow turbulence intensity & $0.0\%$ & $5.0\%$\\
    \end{tabular}
\end{table}

The second simulation case employs more realistic wind conditions to evaluate the potential of Dynamic IPC. These simulations are of a neutral Atmospheric Boundary Layer (ABL) in which the inflow was generated through a so-called precursor simulation. Several properties of both simulation setups are listed in Table~\ref{tab:SOWFA}.

Two different wind farm cases are investigated in these conditions. Firstly, simulations with a single turbine, in which the effects on the turbine and wake are investigated, have been exectuted. Then, a second turbine is added, to assess the gain in energy capture that can be achieved with DIPC. The second turbine is situated 5 rotor diameters ($5D$) behind the upstream turbine, the same axial distance as investigated in \citep{frederik2019}. All these results are presented in Section~\ref{sec:results}.

\section{Control Strategy}\label{sec:approach}

In this section, the Dynamic Individual Pitch Control (DIPC) strategy is further elaborated, as well as the already existing control strategies with which it will be compared. In Section~\ref{sec:sic}, static induction strategies are explained, which includes \textit{greedy} control, where each turbine operates using its individual steady-state optimal settings. These strategies are currently the industry standard, and commonly applied in actual wind farms. They therefore serve as a useful baseline case for cutting edge control concepts such as periodic Dynamic Induction Control (DIC) and the novel DIPC approach. Periodic DIC, as described in \citep{frederik2019}, is shortly covered in Section~\ref{sec:dic}, and Section~\ref{sec:dipc} presents a thorough explanation of the DIPC approach as proposed in this paper. 

\subsection{Static Induction Control}\label{sec:sic}

Static Induction Control (SIC) is a generic term for all induction control strategies that use time-invariant control set-points that depend on the inflow conditions. The most simple static induction wind farm control strategy is to operate all turbines at their individual (static) optimum for power production. This approach is called \emph{greedy control}, as all turbines greedily extract as much power from the wind as possible. As this approach is the simplest and most commonly applied, greedy control is considered the baseline case in this paper.

An alternative approach is to (statically) lower the induction factor, i.e., the in-wake velocity deficit, of upstream turbines such that downstream turbines can increase their power capture. This has for long been the most popular concept in wind farm control research, but recent studies show that the achievable gains with respect to greedy control are minor to non-existent \citep{annoni,Campagnolo2016a,nilsson}. Nonetheless, SIC for power maximization remains of interest to the industry. Hence, it is used as a comparison case in this article to show the potential of DIPC.

\subsection{Periodic Dynamic Induction Control}\label{sec:dic}

A recent research area of interest, as an alternative to SIC, is Dynamic Induction Control (DIC). With this control method, the induction factor of an upstream turbine is varied over time to enhance wake mixing, such that downstream turbines experience higher wind velocities and can subsequently increase their power production. Finding the optimal time-varying induction settings is a very complex control problem \citep{Munters:2017}. A more practical approach is proposed in \citep{Munters:2018}, where sinusoidal input signals on the thrust force $C_T'$ are suggested. This method is called Periodic DIC and will also be used in this paper. It is shown to increase the power production of small wind farms both in simulations \citep{Munters:2018} and in wind tunnel experiments \citep{frederik2019}.

\begin{figure}[b!]
    \centering
    \includegraphics[width=0.55\textwidth]{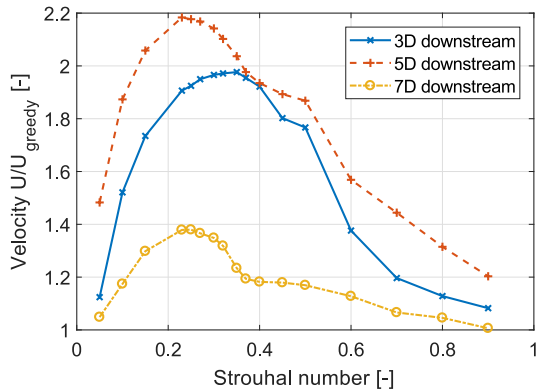}
    \caption{The average wake velocity at $3D$, $5D$ and $7D$ behind an DIC-excited turbine with an pitch amplitude of 4 degrees, for different Strouhal numbers $St$. The results are normalized with respect to the baseline case.}
    \label{fig:gridst}
\end{figure}

In \citep{frederik2019}, for reasons of practical implementation, a periodic excitation is realized by superimposing a low-frequent sinusoidal signal on the static collective pitch angles of the turbine. This approach will also be used in this paper. As the control signal is now confined to a sinusoid, the control parameters are reduced to the amplitude and the frequency of excitation. The frequency is usually characterized in terms of the dimensionless Strouhal number $St$: 
\begin{equation}
    St = \dfrac{f_e D}{U_{\infty}},
    \label{eq:st}
\end{equation}
where $f_e$ is the frequency [Hz], $D$ the rotor diameter [m] and $U_{\infty}$ the inflow velocity [m/s]. As the Strouhal number is dimensionless, it accounts for different turbine sizes or inflow velocities. In the above-mentioned references, an optimal Strouhal number of $St \approx 0.25$ is found. For the DTU 10MW turbine \citep{DTU10MW}, with an inflow velocity of 9~m/s, an excitation frequency of $f_e = 0.0126$~Hz is found. To verify this optimal frequency, an extensive evaluation is performed in SOWFA. A single 10MW wind turbine is placed in laminar flow conditions (see Table~\ref{tab:SOWFA}), and the velocity is measured at integer multiples of the rotor diameters $D$ behind the turbine. The results are presented in Figure~\ref{fig:gridst} and show that for a distance $\geq 5D$, the optimum is indeed around $St = 0.25$. As a physical explanation for the optimal frequency is not yet investigated, this excitation frequency was used in the simulations presented here. To take into account the effect of different excitation amplitudes, two different DIC cases will be considered: a low amplitude case with a collective pitch amplitude of $2.5^{\circ}$ and a high amplitude case of $4^{\circ}$, respectively.

\subsection{Dynamic Individual Pitch Control}\label{sec:dipc}

In this section, the novel Dynamic Individual Pitch Control (DIPC) approach will be described. The goal of this approach is to enhance wake mixing analogous to DIC, but without the large fluctuations in thrust and power. To achieve this, the individual pitch angles are altered in such a way that the wake behind the excited turbine is manipulated dynamically.

Fundamentally, DIPC works as follows. The individual blade pitch angles of the turbine can be used to generate a directional moment on the rotor. Consequently, the direction of the force vector exerted on the airflow can be manipulated. With DIPC, the direction of this force vector is slowly varied, thereby continuously changing the direction of the wake. This is expected to increase wake mixing without significant variations in the magnitude of the rotor thrust force.

A directional thrust force can be accomplished by implementing the Multi-Blade Coordinate (MBC) transformation~\citep{mbc}. This transformation decouples -- or stated differently: \textit{projects} -- the blade loads in a non-rotating reference frame. As a result, the measured out-of-plane blade root bending moments $M(t)\in\mathbb{R}^3$ are projected onto a non-rotating reference frame. For a three-bladed turbine, the MBC transformation is given as:

\begin{align}
\left[ \begin{array}{c} M_{0}(t) \\ M_{{\textrm{tilt}}}(t) \\ M_{{\textrm{yaw}}}(t) \end{array} \right] &= \mathbf{T}(\psi) \underbrace{\left[ \begin{array}{c} M_{1}(t) \\ M_{2}(t) \\ M_{3}(t) \end{array} \right]}_{M(t)}\label{eq:mbc},
\end{align}
with
\begin{align}
\nonumber\mathbf{T}(\psi) &= \frac{2}{3}\left[ \begin{array}{c c c} 0.5 & 0.5 & 0.5 \\ \cos{\left(\psi_1\right)} & \cos{\left(\psi_2\right)} & \cos{\left(\psi_3\right)} \\ \sin{\left(\psi_1\right)} & \sin{\left(\psi_2\right)} & \sin{\left(\psi_3\right)} \\ \end{array} \right],
\end{align}

\begin{figure}[b]
    \centering
    \includegraphics[width=\textwidth]{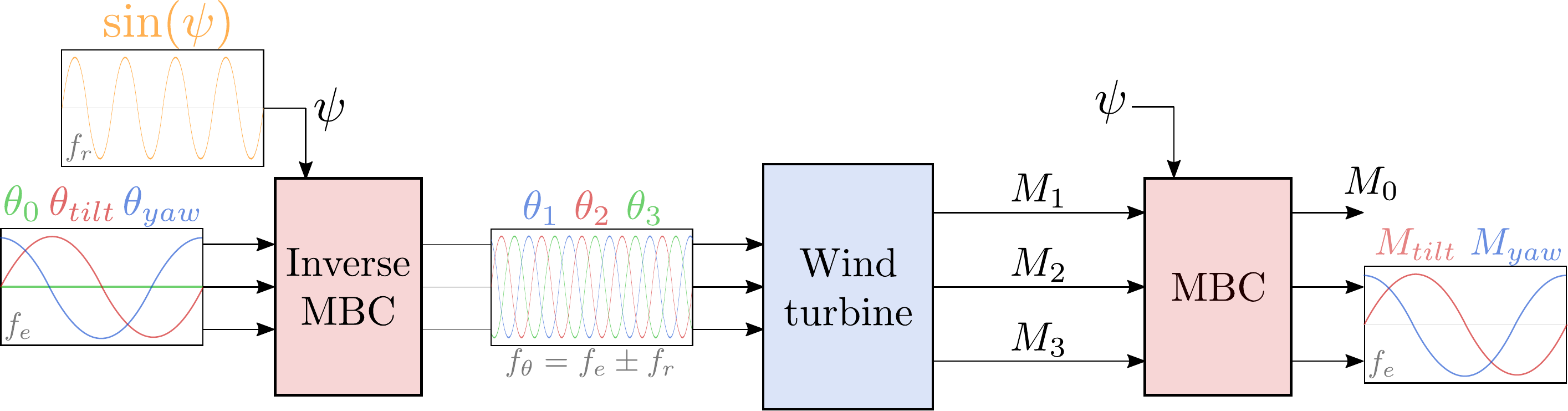}
    \caption{A schematic representation of how the MBC transformation is used to achieve periodic yaw and tilt moments on the turbine. Note that the pitch frequency $f_{\theta}$ is slightly different than the rotation frequency $f_r$ due to excitation frequency $f_e$.}
    \label{fig:mbc_ol}
\end{figure}

where $\psi_b$ is the azimuth angle for blade $b$, with $\psi = 0^{\circ}$ indicating the vertical upright position. The collective mode $M_0$ represents the cumulative out-of-plane rotor moment, and $M_{\textrm{tilt}}$ and $M_{\textrm{yaw}}$ represent the fixed-frame and azimuth-independent tilt- and yaw-moments, respectively. 


In a similar fashion, the inverse MBC transformation can be used to obtain implementable pitch angles based on the fixed-frame collective, tilt and yaw pitch signals, $\theta_0$, $\theta_{\textrm{tilt}}$ and $\theta_{\textrm{yaw}}$, respectively:

\begin{align}
\underbrace{\left[ \begin{array}{c} {\theta}_{1}(t) \\ {\theta}_{2}(t) \\ {\theta}_{3}(t) \end{array} \right]}_{{\theta}(t)} &= \mathbf{T}^{-1}(\psi) \left[ \begin{array}{c} \theta_{0}(t) \\ \theta_{\textrm{tilt}}(t) \\ \theta_{\textrm{yaw}}(t) \end{array} \right],	
\label{eq:invmbc}
\end{align}
with
\begin{align}
\nonumber\mathbf{T}^{-1}(\psi) &= \left[ \begin{array}{ccc} 1 & \cos{\left(\psi_1\right)} & \sin{\left(\psi_1\right)} \\ 1 & \cos{\left(\psi_2\right)} & \sin{\left(\psi_2\right)} \\ 1 & \cos{\left(\psi_3\right)} & \sin{\left(\psi_3\right)} \end{array} \right].
\end{align}

The concept of DIPC is to achieve a dynamically varying tilt and/or yaw moment, such that the wake of the turbine is manipulated in vertical and/or horizontal direction, respectively, over time. To give a proof of concept, a simple feedforward strategy is implemented, where a sinusoidal excitation is superimposed on $\theta_{\textrm{tilt}}$ and $\theta_{\textrm{yaw}}$, as shown in Figure~\ref{fig:mbc_ol}. The excitation frequency of $\theta_{\textrm{tilt}}$ and $\theta_{\textrm{yaw}}$ is chosen to be identical to the DIC case, i.e., $St = 0.25$. Note once more that this is a low-frequent excitation, typically in the range of 10 times slower than the rotational frequency $f_r$.  It will be shown later that the resulting tilt and yaw moments are indeed sinusoidal with frequency $f_e$.

When the tilt and yaw pitch angles inserted into the inverse MBC transformation are constant over time, the resulting pitch angles ${\theta}(t)$ will behave sinusoidally with frequency $f_r$. However, when $\theta_0 = 0$ and the tilt and yaw pitch angles are sinusoidal (with frequency $f_e$), as depicted in Figures~\ref{fig:mbc_ol}, this leads to a slightly altered frequency of ${\theta}(t)$. Using \eqref{eq:invmbc}, it can be deduced that:

\begin{align*}
    \theta_b(t) & = \left[\begin{array}{ccc} 1 & \cos(\psi_b) & \sin(\psi_b)\end{array}\right]\left[\begin{array}{c}\theta_0(t) \\ \theta_{{\textrm{tilt}}}(t) \\ \theta_{{\textrm{yaw}}}(t)\end{array}\right] \\
    & = \theta_0 + \cos(\omega_r t+\psi_{0,b})\theta_{\textrm{tilt}}(t) + \sin(\omega_r t+\psi_{0,b})\theta_{\textrm{yaw}}(t)\\
    & = \cos(\omega_r t+\psi_{0,b})\sin(\omega_e t) + \sin(\omega_r t+\psi_{0,b})\cos(\omega_e t) \\
    & = \sin\left[(\omega_r+\omega_e)t+\psi_{0,b}\right],
\end{align*}

where $\omega_r$ is the rotational velocity [rad/s], and $\omega_e = 2\pi f_e$ [rad/s]. Assuming that $\omega_r$ is constant over time, $\psi_b(t) = \omega_r t + \psi_{0,b}$ with $\psi_{0,b}$ the azimuth angle of blade $b$ at $t=0$. Since the excitation frequency is very low (i.e., $\omega_e \ll \omega_r$) the frequency of the resulting sinusoid, $f_{\theta}$, differs only slightly from the rotational frequency $f_r$.

\begin{figure}[t!]
     \centering
     \begin{subfigure}{0.23\textwidth}
     \includegraphics[width=\textwidth]{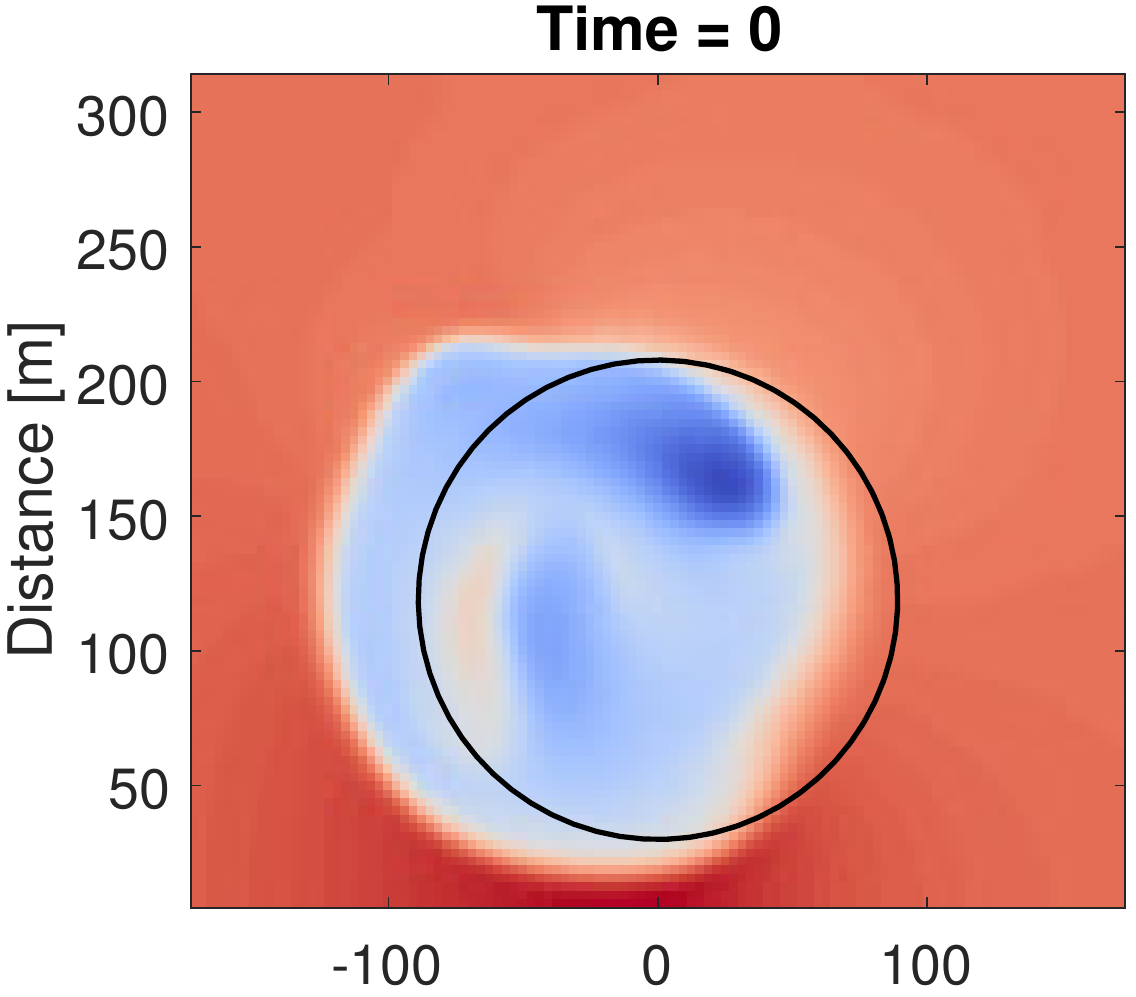}
     \end{subfigure}
     ~
     \begin{subfigure}{0.23\textwidth}
     \includegraphics[width=\textwidth]{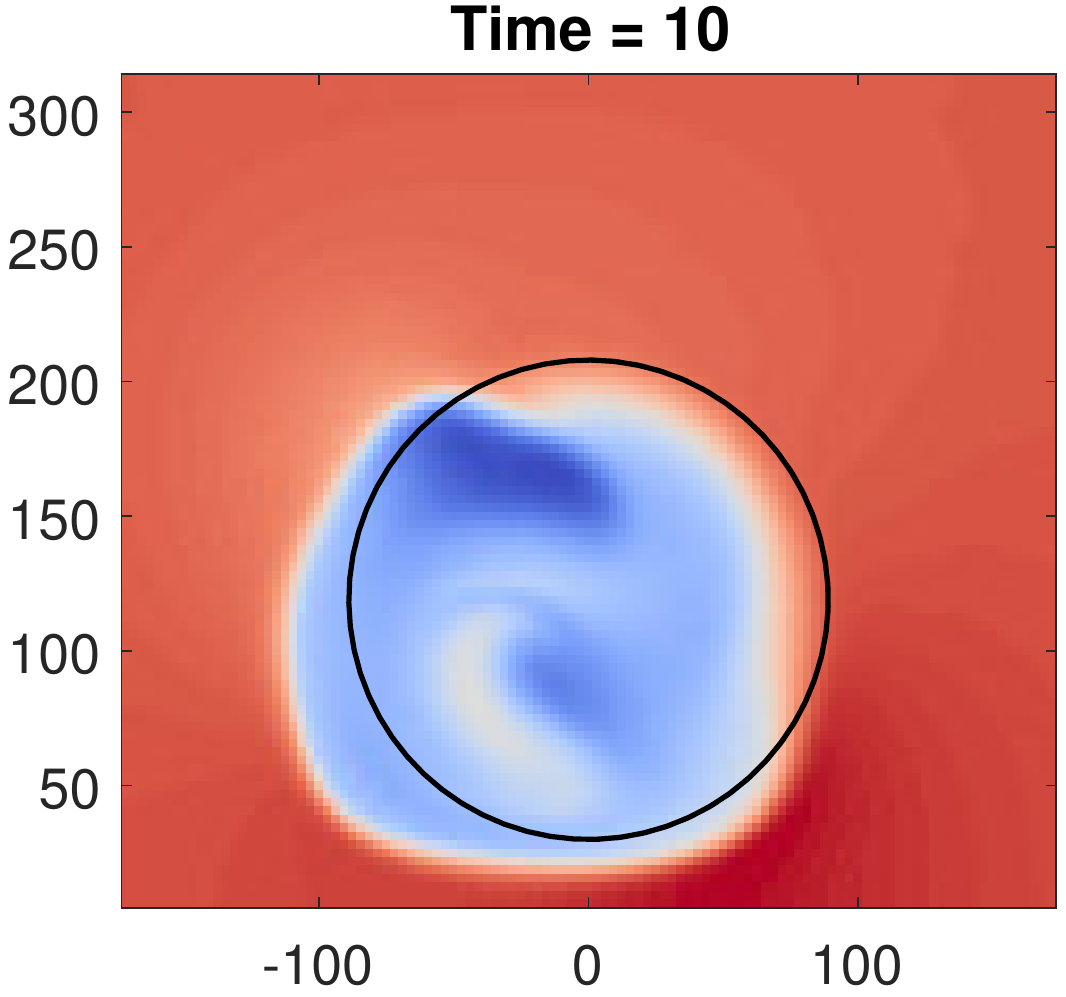}
     \end{subfigure}
     ~
     \begin{subfigure}{0.23\textwidth}
     \includegraphics[width=\textwidth]{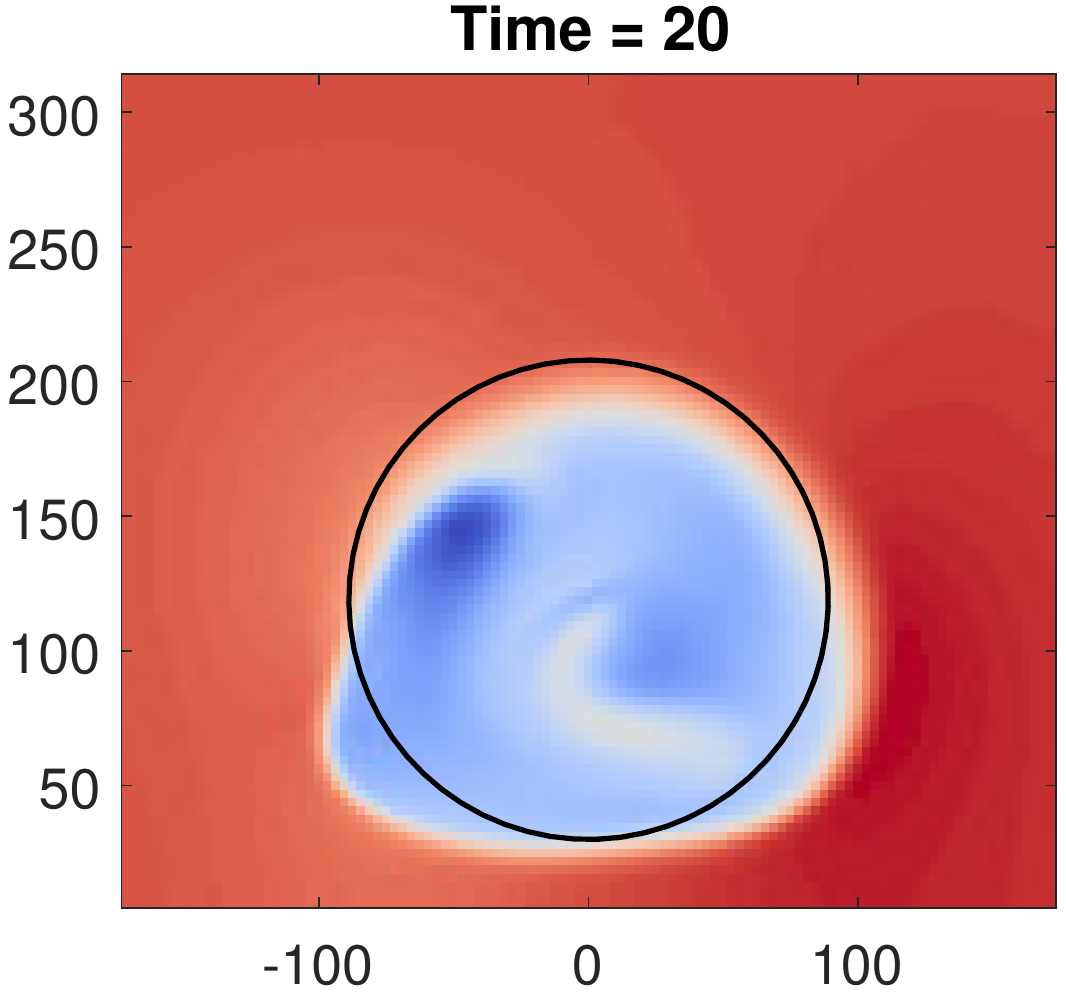}
     \end{subfigure}
     ~
     \begin{subfigure}{0.23\textwidth}
     \includegraphics[width=\textwidth]{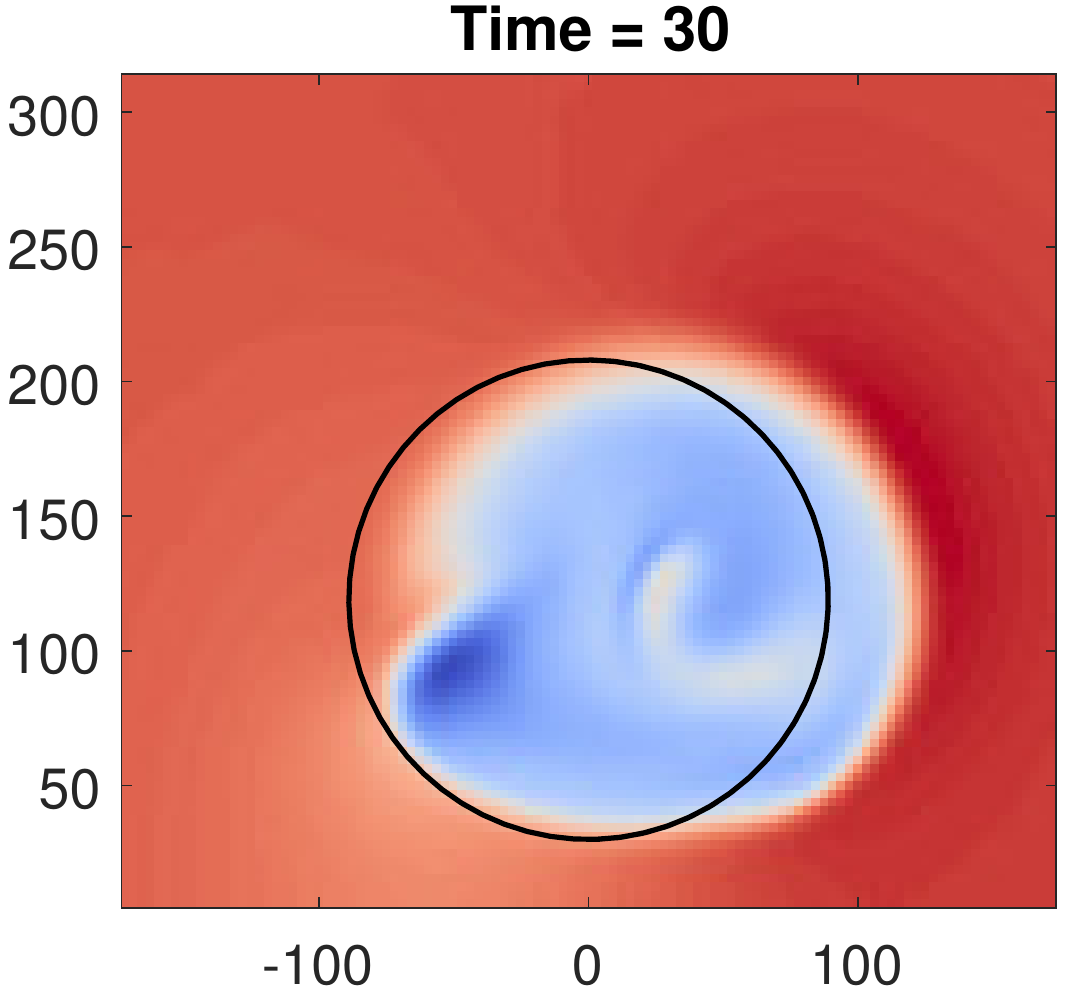}
     \end{subfigure}
    
     \begin{subfigure}{0.23\textwidth}
     \includegraphics[width=\textwidth]{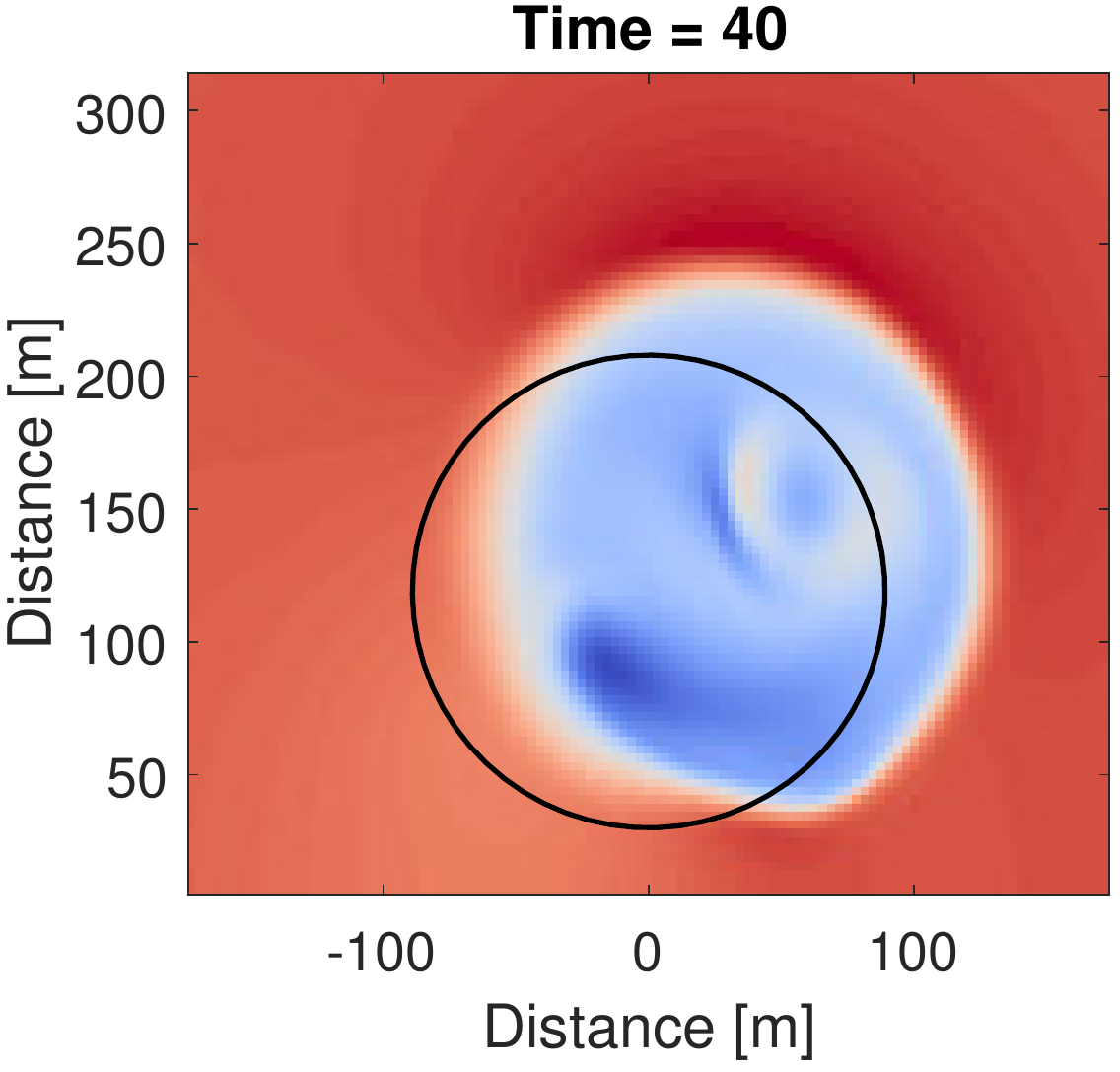}
     \end{subfigure}
     ~
     \begin{subfigure}{0.23\textwidth}
     \includegraphics[width=\textwidth]{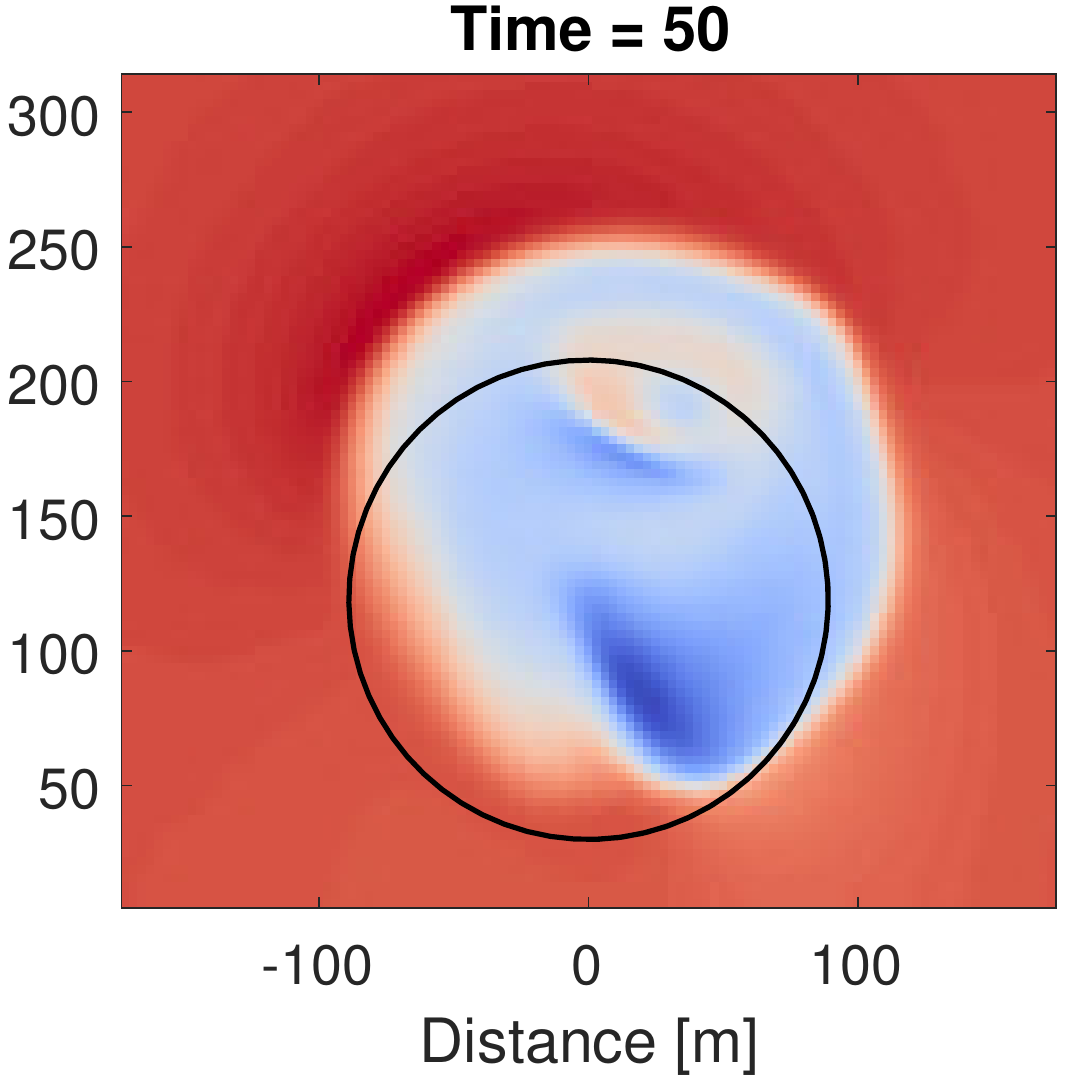}
     \end{subfigure}
     ~
     \begin{subfigure}{0.23\textwidth}
     \includegraphics[width=\textwidth]{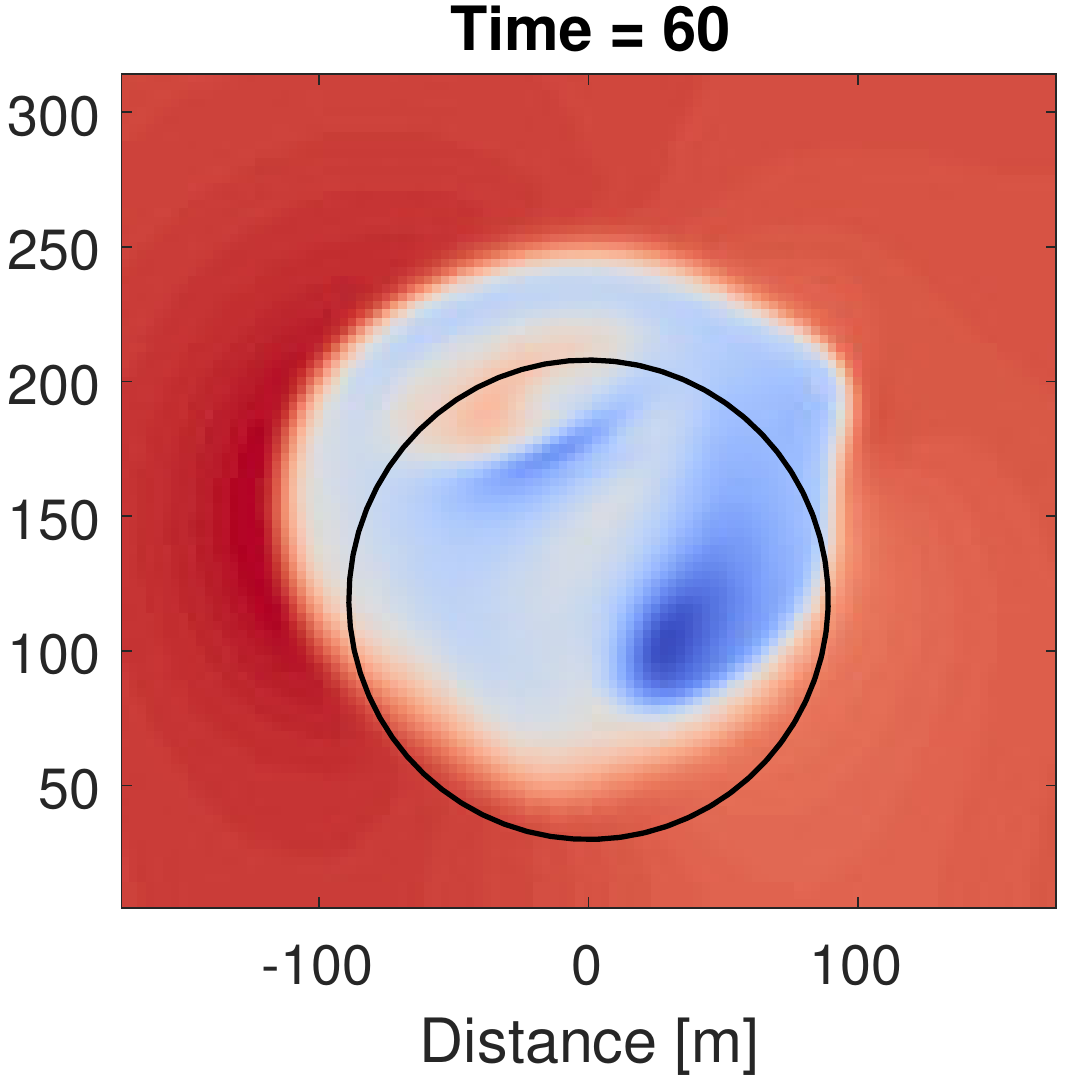}
     \end{subfigure}
     ~
     \begin{subfigure}{0.23\textwidth}
     \includegraphics[width=\textwidth]{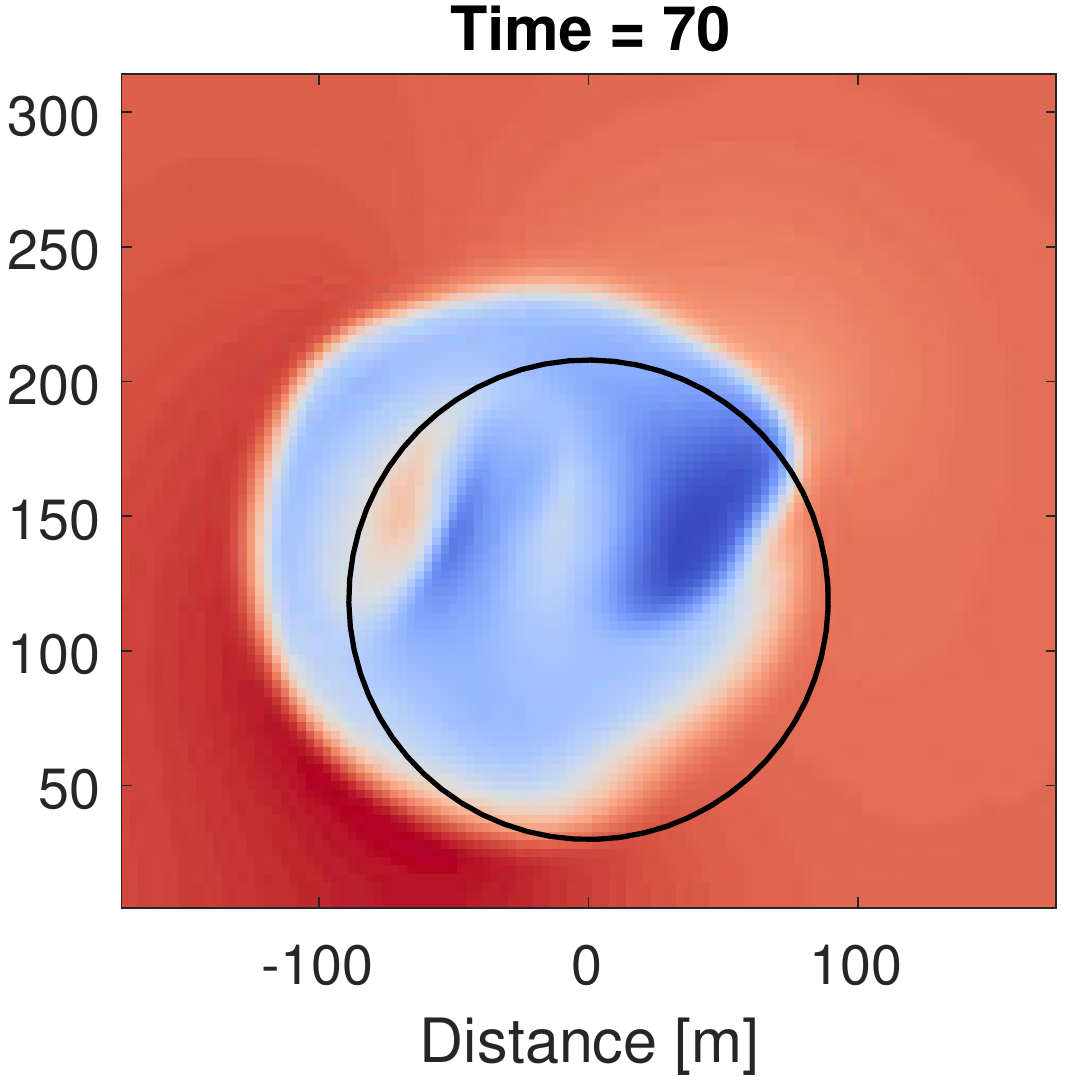}
     \end{subfigure}
     \caption{A wake as measured at 3$D$ behind the turbine at different time instances when the signals for $\theta_{tilt}$ and $\theta_{yaw}$ as displayed in Figure~\ref{fig:mbc_ol} are applied. Obtained using uniform inflow simulations in SOWFA.}
     \label{fig:wakehelix}
 \end{figure}

In Figure~\ref{fig:mbc_ol}, a shift of 90 degrees between the yaw moment and the tilt moment is depicted. As a result, the tilt moment is maximal when the yaw moment is zero, and vice versa. Using the uniform simulation setup in SOWFA, the resulting wake location over time can be visualized. Figure~\ref{fig:wakehelix} shows this wake at eight instances during one excitation period $T_e = D/(St U_\infty)\approx 80$~s. It can be observed here that this DIPC strategy results in a wake that makes a (counterclockwise) circular motion. This motion can be considered forced wake meandering, and is expected to lead to increased wake mixing.

  \begin{figure}[t!]
     \centering
     \includegraphics[width=\textwidth]{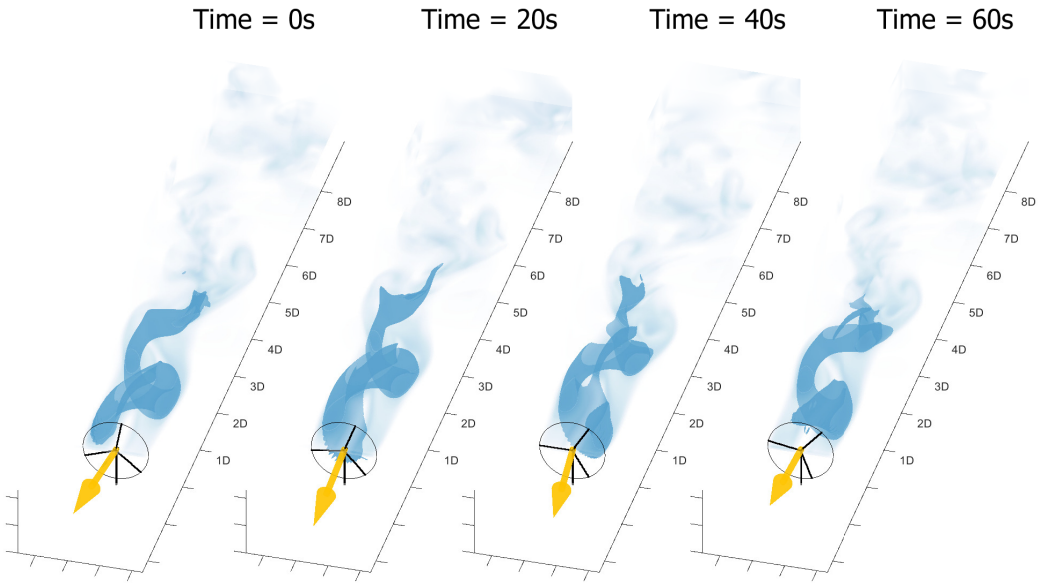}
     \caption{Wake propagation for different time instances when the helix strategy is applied. The counterclockwise rotation of the wake can be seen and the near wake clearly exhibits the helix shape that the approach is named after. The yellow arrow represents the vector of the thrust applied on the flow.}
     \label{fig:helixshape}
 \end{figure}

\clearpage

Figure~\ref{fig:wakehelix} displays the wake for a phase delay of 90 degrees between the tilt and yaw pitch angle, leading to a counterclockwise motion of this wake. It is also possible to force the wake into a clockwise motion by applying a phase delay of 270 degrees. In that case, the resulting pitch frequency will be slightly \textit{lower} than the rotation frequency:

\begin{align*}
    \theta_b(t) & = \cos(\omega_b t)\sin(\omega_e t) - \sin(\omega_b t)\cos(\omega_e t) \\
    & = -\sin\left[(\omega_b-\omega_e)t\right].
\end{align*}

As the resulting wake propagates through space in a helical fashion, this specific approach is called the \textit{helix strategy}, respectively in counterclockwise (CCW) or clockwise (CW) direction. This helical wake propagation is illustrated in Figure~\ref{fig:helixshape}.

\begin{figure}
    \centering
    \includegraphics[width=\textwidth]{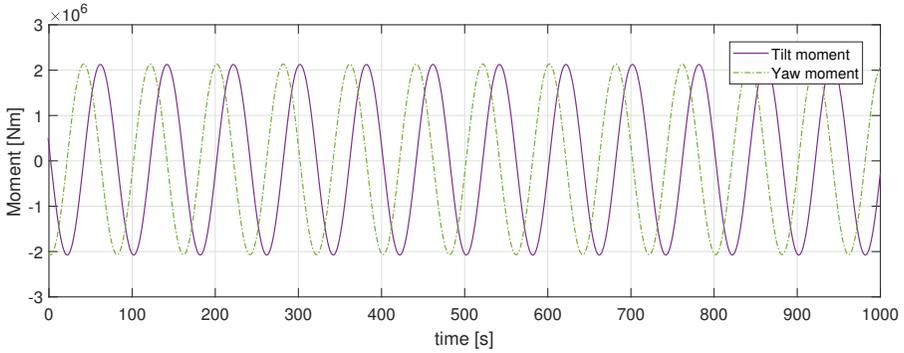}
    \caption{The tilt and yaw moments from a turbine operating with the CCW helix approach. Obtained using uniform inflow conditions in SOWFA.}
    \label{fig:tiltyawmoments}
\end{figure}

Earlier in this section, the claim was made that a sinusoidal tilt and yaw moment can be achieved by simply applying a sinusoidal tilt and yaw angle. To confirm that this is indeed the case, Figure~\ref{fig:tiltyawmoments} shows the tilt and yaw moment for the CCW helix strategy. These moments were obtained using the out-of-plane root bending moments on the individual blades as obtained from SOWFA, subsequently projected onto the non-rotating frame using the MBC transformation~\eqref{eq:mbc}. Afterwards, a low-pass filter was applied to account for high frequency noise. Note that the amplitude of both signals is identical, and that a phase offset of 90 degrees can indeed be observed.

\section{Results}\label{sec:results}

In this section, the results obtained from the SOWFA simulations with turbulent inflow, as described in Section~\ref{sec:sowfa}, are presented. The helix approach is compared to the baseline greedy control case, as well as with Static Induction Control (SIC) and Dynamic Induction Control (DIC). First, simulations with a single turbine are evaluated. These simulations allow for the investigation of the helix approach on the excited turbine and on the wake behind this turbine. Afterwards, a second turbine is placed in the wake, 5$D$ behind the first turbine, to study the effect of DIPC on this downstream machine and on the power production of this small 2-turbine wind farm.

 \begin{figure}[b!]
     \centering
     \includegraphics[width=\textwidth]{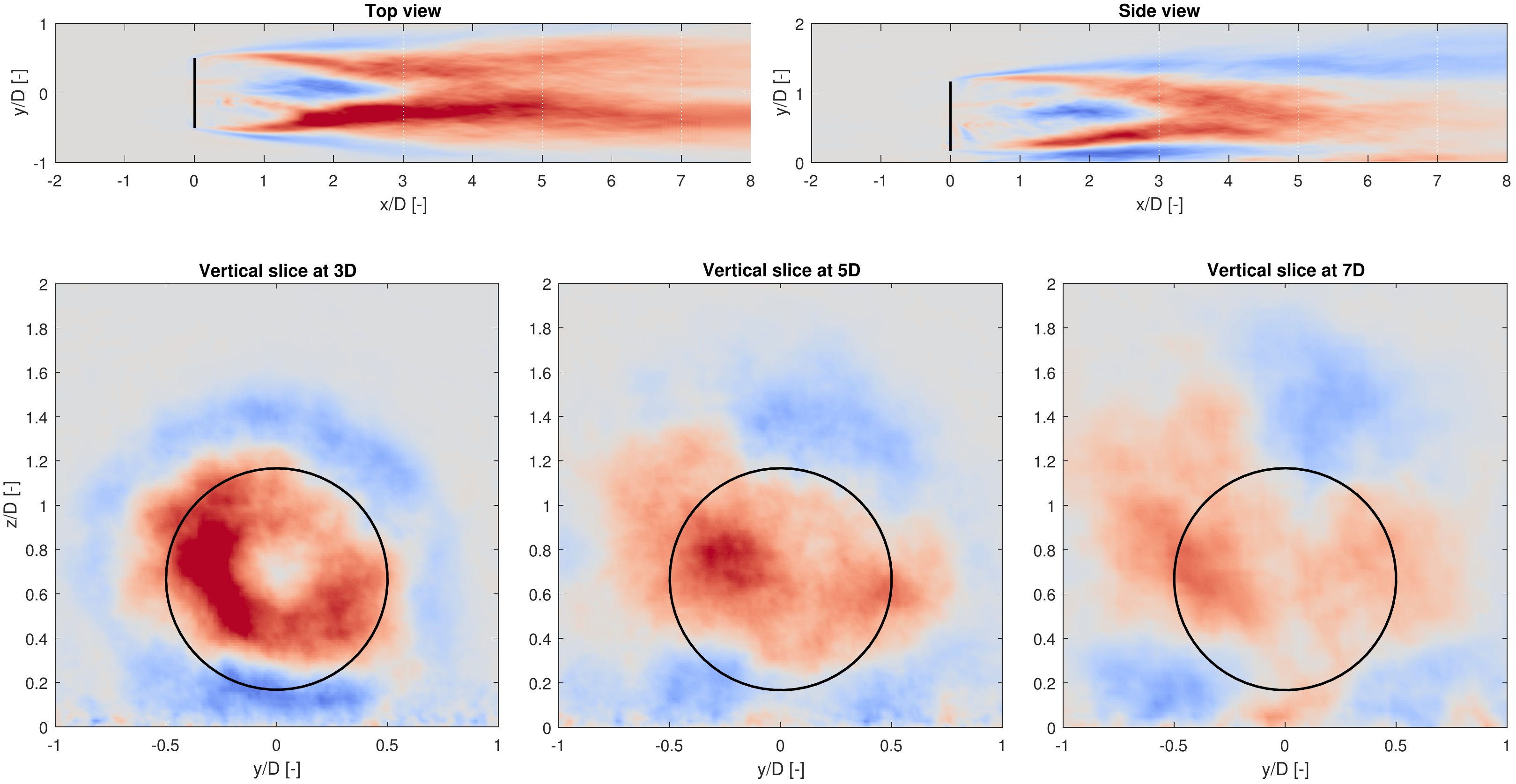}
     \caption{The mean wind speed in a wake with respect to the baseline case when DIC is applied (case 7). The turbine location is indicated in black. The top figures give a top and side view of the flow, and the bottom figures show vertical slices at different distances behind the excited turbine. The red areas indicate that DIC increases the wind velocity in the wake significantly, while blue areas indicate where the wind speed is decreased.}
     \label{fig:flow_dic}
 \end{figure}

\subsection{Single turbine}

For the single turbine case, a total of 9 different simulations have been carried out. A comparison between cases will be made based on both the performance of the turbine and the energy available in the wake. The simulation cases are specified below:

\begin{enumerate}
    \item \textit{Baseline case}: static greedy control. All other cases will be normalized with respect to this case;
        \item \textit{Static Induction Control (SIC), 1$^\circ$}: SIC where the collective pitch angles are derated with 1$^\circ$;
    \item \textit{SIC, 2$^\circ$}: Same as case 8, but with the pitch angles derated 2$^\circ$;
    \item \textit{Dynamic Induction Control (DIC), 2.5$^\circ$}: DIC where the collective pitch angles are excited sinusoidally with an amplitude of $2.5^\circ$;
    \item \textit{Counterclockwise (CCW) helix, 2.5$^\circ$}: the helix approach with a phase offset between tilt and yaw moments of 90 degrees (as shown in Figure~\ref{fig:mbc_ol}). This results in a wake that rotates in counterclockwise direction. The amplitude of the tilt and yaw angles is chosen such that the variation of the implemented pitch angles has an amplitude of $2.5^\circ$;
    \item \textit{Clockwise (CW) helix, 2.5$^\circ$}: the helix approach with a phase offset between tilt and yaw moments of 270$^\circ$. This results in a wake rotating in clockwise direction;
    \item \textit{DIC, 4$^\circ$}: Same as case 2, but with an amplitude of 4$^\circ$;
    \item \textit{CCW helix, 4$^\circ$}: Same as case 3, but with an amplitude of 4$^\circ$;
    \item \textit{CW helix, 4$^\circ$}: Same as case 4, but with an amplitude of 4$^\circ$.
\end{enumerate}

\begin{figure}[t!]
     \centering
     \includegraphics[width=\textwidth]{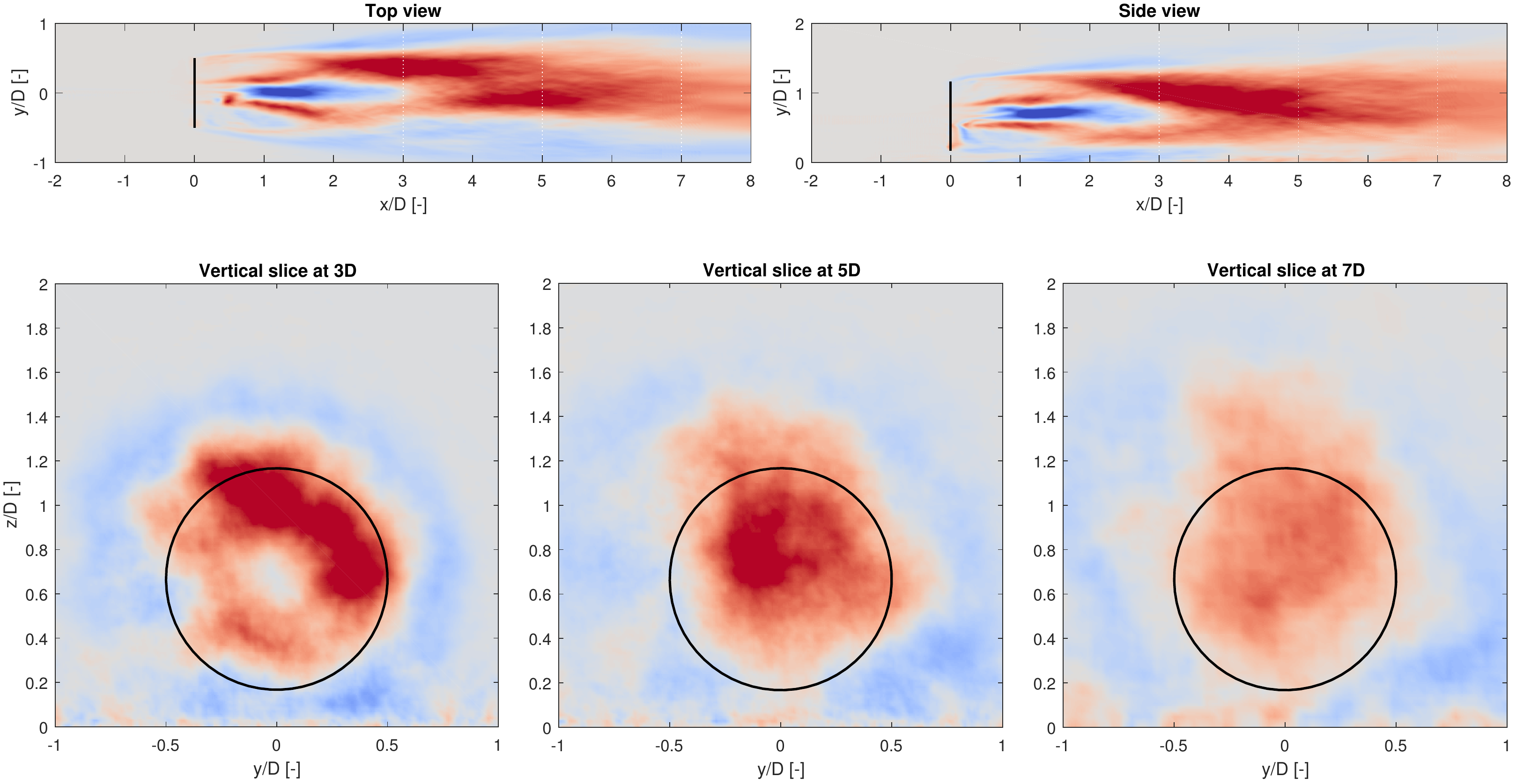}
     \caption{The mean wind speed in a wake with respect to the baseline case when CCW helix is applied (case 8), similar to Figure~\ref{fig:flow_dic}.}
     \label{fig:flow_ccw}
 \end{figure}

\begin{figure}[t!]
     \centering
     \includegraphics[width=\textwidth]{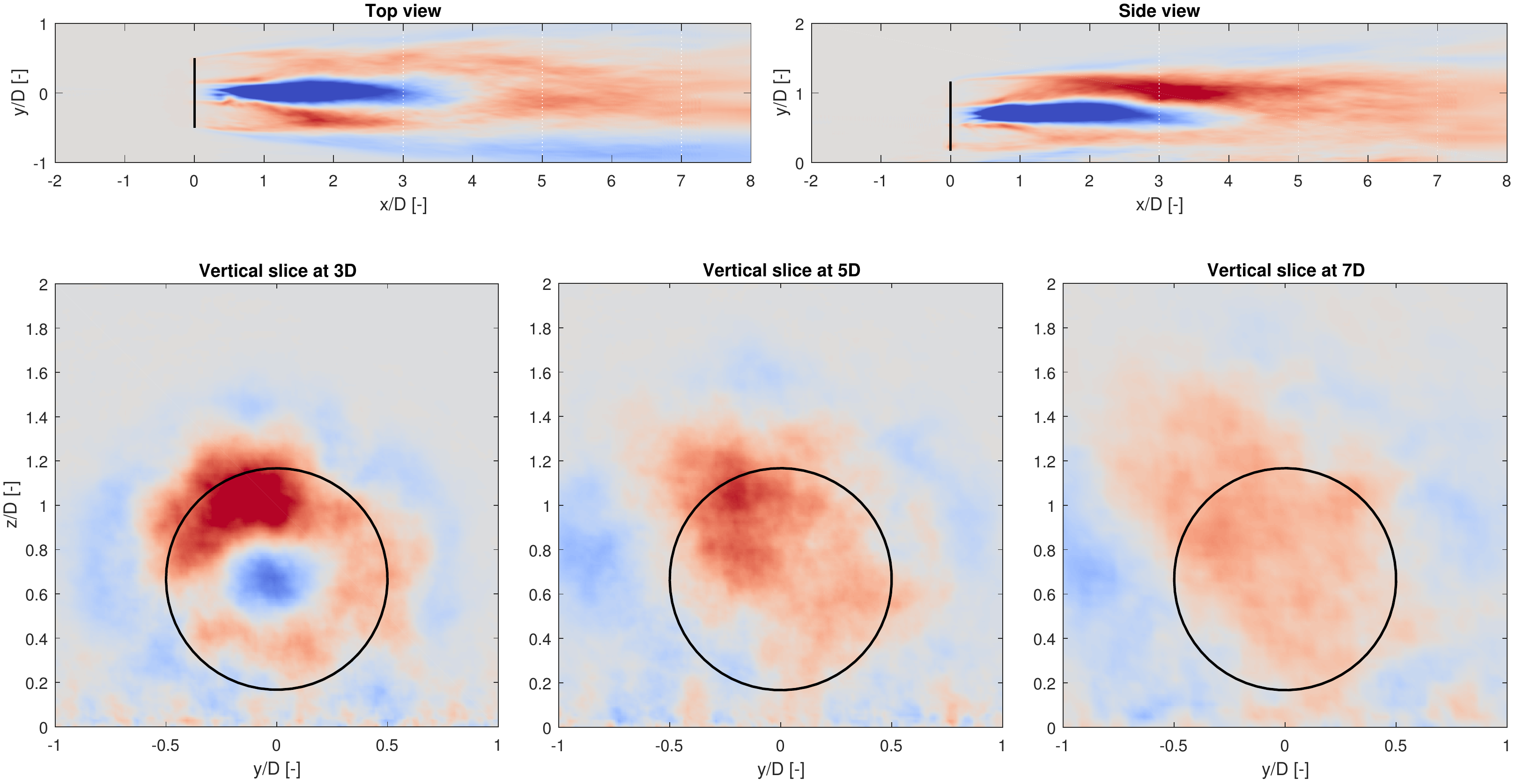}
     \caption{The mean wind speed in a wake with respect to the baseline case when CW helix is applied (case 9), similar to Figures~\ref{fig:flow_dic}~and~\ref{fig:flow_ccw}.}
     \label{fig:flow_cw}
 \end{figure}

First of all, the effect of the helix approach on the wake is investigated. For this purpose, the mean wind velocity behind the excited turbine is visualized with respect to the baseline case. The resulting figures show how the applied control algorithms change the wake properties. Figure~\ref{fig:flow_dic} shows this mean velocity disparity with respect to the baseline case for case 7 (DIC, 4$^\circ$). Different cross-sections of the flow field are depicted here to show the effect of DIC on the average wake velocity. Figure~\ref{fig:flow_ccw} depicts the same cross-sections for the case 8 (CCW helix, 4$^\circ$) and Figure~\ref{fig:flow_cw} for case 9 (CW helix, 4$^\circ$). Remember that, as mentioned in Section~\ref{sec:dipc}, the optimal amplitude and frequency for the helix approach are as of yet unknown. The results presented here should therefore be considered a proof of concept for this approach, not an upper limit of its potential.

Based on these figures, a number of conclusions are drawn. First of all, it is clear that all three strategies successfully increase the average wind velocity in the wake. DIC and CCW helix seem to be equally effective at 3$D$, while the helix approach performs increasingly well further downstream. In general, the CW helix appears to be less effective than the CCW helix. Figure~\ref{fig:flow_cw} reveals that the lower performance of the CW approach is caused by the lower velocity in the center of the wake, which is considerably more distinct than in Figure~\ref{fig:flow_ccw}.

The average kinetic energy increase in the wake at 5$D$ behind the turbine is 23.8$\%$ for DIC, 36.7$\%$ for CCW helix and 19.3$\%$ for CW helix. This indicates that the power increase that can be expected of a second, waked turbine when the CCW helix is applied will be higher than in the DIC case.

\begin{figure}[t!]
    \centering
    \includegraphics[width=\textwidth]{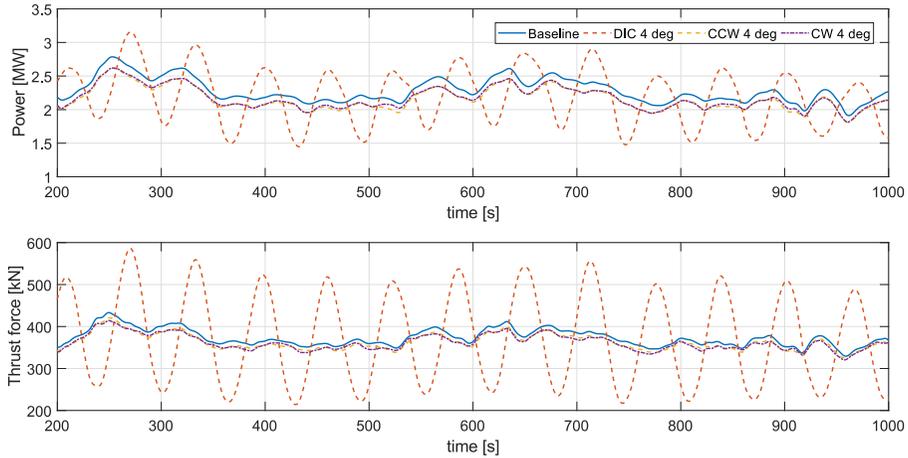}
    \caption{The power (top) and thrust (bottom) signals of the wind turbine for the baseline, DIC, CCW helix and CW helix case.}
    \label{fig:powerthrust}
\end{figure}

The mean wake velocities are nonetheless not the most significant difference between DIC and the helix approach. The main advantage of the helix approach becomes clear when the power and thrust signals of the excited turbine are examined, as shown in Figure~\ref{fig:powerthrust}. These plots shows that, as expected, DIC results in large variations of both the power production and the thrust force. Both helix approach simulations show no such variations: the power and thrust are in both cases very similar to the baseline case, although slightly lower. This is also confirmed when the variance of these signals is calculated. When DIC is applied, the variance of the power and the thrust increases -- compared to the baseline case -- with 80\% and 583\%, respectively. With the helix approach, on the other hand, the variance of these signals stays more or less the same with respect to the baseline case. 

This significant improvement with regards to the thrust and power variations does not come completely free of charge. Since Individual Pitch Control is used for the helix method, the pitch rate, and subsequently the actuator duty cycle, is higher than with DIC. As visualized in Figure~\ref{fig:mbc_ol}, the frequency of the pitch signal is determined by the rotational frequency $f_r \approx 0.12$\,Hz, slightly altered by the excitation frequency. The pitch signals in DIC, on the other hand, have a much lower frequency of $f_e \approx 0.0126$\,Hz, resulting in a very low average pitch rate variation of 0.08\,$^{\circ}$/min. As a consequence of the higher pitch frequency, the pitch rate variance of the helix approach with a 4 degree amplitude is 12.5$^{\circ}$/min and 8.1$^{\circ}$/min for the CCW and CW direction, respectively. Note that although this is significantly higher than with DIC, such a pitch rate should not be considered unreasonably high. In fact, the pitch rate is comparable to that used in load alleviating IPC strategies such as \citep{bossanyi2003individual,bossanyi2005further}.

All the results mentioned above, both in terms of turbine performance and wake recovery, are summarized in Table~\ref{tab:1T_turb}. This table includes the results obtained for the cases with SIC and with a smaller pitch amplitude of 2.5$^{\circ}$. As expected, the lower amplitude has less effect on both the excited turbine and the wake recovery. Apart from that, no significant discrepancies are found between the 2.5$^{\circ}$ and 4$^{\circ}$ cases. The SIC results show that, in general, the power lost at the upstream turbine is comparable, while the energy gained in the wake is lower than with the CCW helix approach. Even more so than DIC, SIC seems to be less effective at larger downstream distances. It can therefore be concluded that the helix approach is more effective in increasing the potential energy capture of a wind farm than SIC. 

\begin{table}[t!]
    \centering
    \setlength{\tabcolsep}{5pt} 
    \caption{Turbulent inflow, single turbine results. All but the pitch rate are relative results with respect to the baseline case.}
    \begin{tabular}{l|c|c|c|c|c|c|c|c}
         & \rotatebox[origin=c]{90}{\textbf{Static 1$^\circ$}} & \rotatebox[origin=c]{90}{\textbf{Static 2$^\circ$}} & \rotatebox[origin=c]{90}{\textbf{DIC 2.5$^\circ$}} &  \rotatebox[origin=c]{90}{\textbf{CCW Helix 2.5$^\circ$}} & \rotatebox[origin=c]{90}{\textbf{CW Helix 2.5$^\circ$}} & \rotatebox[origin=c]{90}{\textbf{DIC 4$^\circ$}} & \rotatebox[origin=c]{90}{\textbf{CCW Helix 4$^\circ$}} & \rotatebox[origin=c]{90}{\textbf{CW Helix 4$^\circ$}}  \\ \hline
        \textbf{Power} & \color{red}{-1.0\%} & \color{red}{-3.1\%} & \color{red}{-1.1\%} & \color{red}{-1.1\%} & \color{red}{-1.0\%} & \color{red}{-2.8\%}  & \color{red}{-2.8\%} & \color{red}{-2.6\%} \\
        \textbf{Variation of power} & -2.2\% & -5.8\% & \color{red}{+79.5\%} & -3.5\% & -1.5\% & \color{red}{+194.1\%} & -7.9\% & -5.2\%  \\
        \textbf{Variation of thrust} & -11.2\% & --22.1\% & \color{red}{+583.2\%} & -1.5\% & \color{red}{+1.2\%} & \color{red}{+1423\%}  & -3.7\% & \color{red}{+0.3\%} \\
        \textbf{Energy at 3$D$} & +14.1\% & +31.7\% & +20.3\% & +20.5\% & +7.2\% & +42.6\%  & +47.4\% & +21.9\% \\
        \textbf{Energy at 5$D$} & +3.7\% & +8.3\% & +13.3\% & +16.6\% & +6.5\% & +23.8\%  & +36.7\% & +19.3\% \\
        \textbf{Energy at 7$D$} & +2.0\% & +3.7\% & +7.3\% & +12.4\% & +5.5\% & +13.4\% & +25.6\% & +14.7\% \\
        \textbf{Pitch variation} [$^{\circ}$/min] & 0 & 0 & 0.08 & \color{red}{4.94} & \color{red}{3.22} & 0.20 & \color{red}{12.52} & \color{red}{8.13}
    \end{tabular}
    \label{tab:1T_turb}
\end{table}

\subsection{Two-turbine wind farm}

In this section, the performance of a two-turbine wind farm is discussed. The same cases of the single turbine simulations are used, but a second turbine is now placed $5D$ behind the first turbine. In all cases, the second turbine operates at its static optimum, i.e., the different control strategies are only implemented on the upstream machine.

\begin{figure}[b!]
    \centering
    \includegraphics[width=\textwidth]{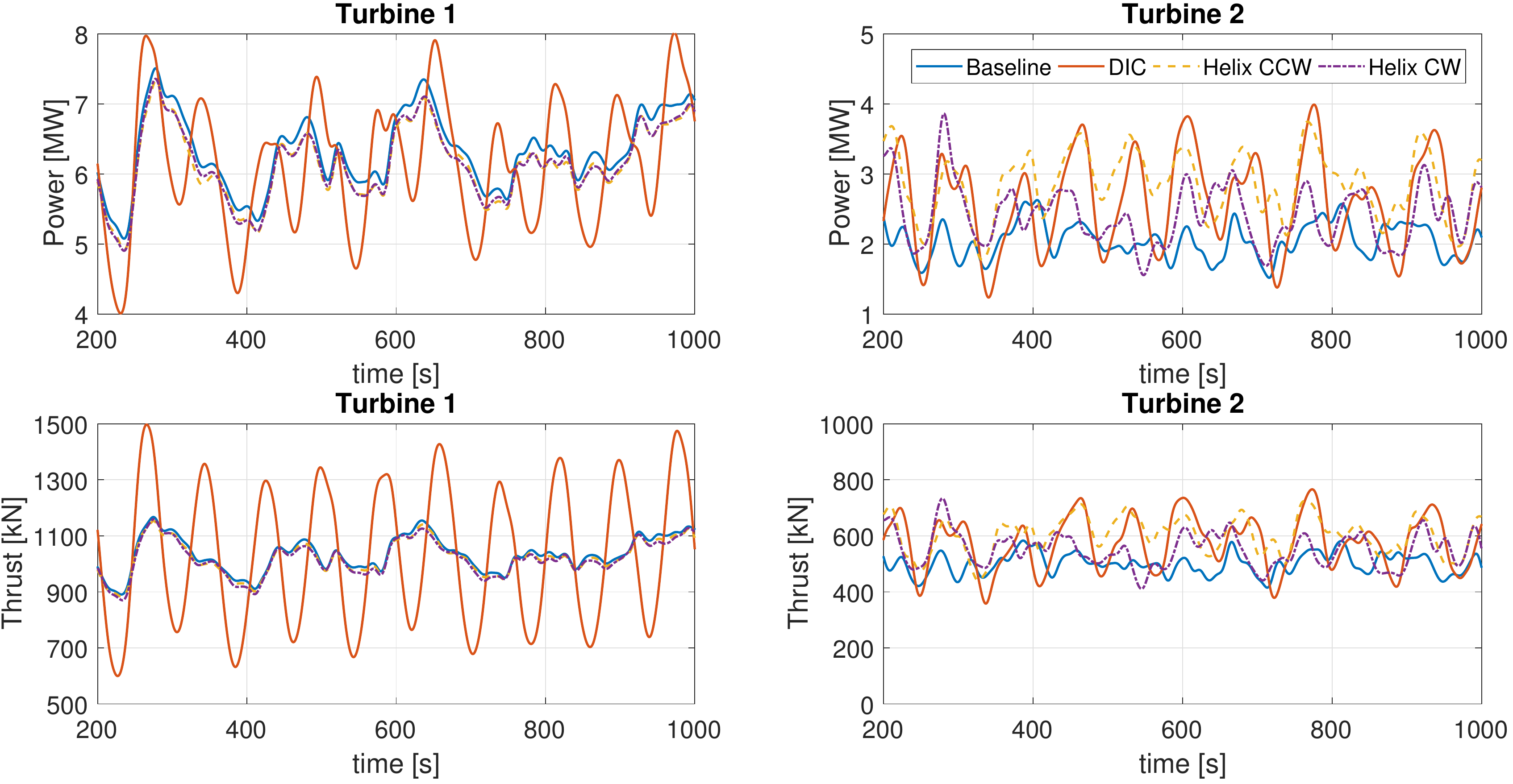}
    \caption{The power (top) and thrust (bottom) signals of turbines 1 (left) and 2 (right) for cases 1, 5, 6 and 7. The variations in power and thrust associated with DIC are not present with the helix approach. As a result, the power and thrust variations at the downstream turbine are also significantly lower.}
    \label{fig:pt2t}
\end{figure}

The results that are presented here, focus again on the cases with a pitch amplitude of 4 degrees. The power and thrust signals of both turbines in these simulations are shown in Figure~\ref{fig:pt2t}. As expected, the power production of turbine 2 when DIC is implemented on turbine 1 is slightly higher than with the helix strategies. However, the plot also shows that DIC not only increases the variations in power and thrust of the excited turbine, but also of the downstream turbine. This effect is significantly less pronounced for the helix strategies.

\begin{figure}[t!]
    \centering
    \includegraphics[width=0.9\textwidth]{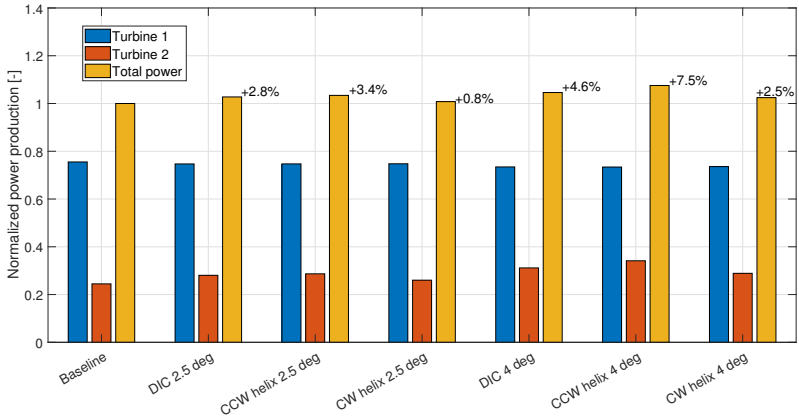}
    \caption{Power production of the two-turbine wind farm for different control strategies, showing the limited power loss at turbine 1 with all methods. The power gain at turbine 2 results in a farm-wide increase in power production.}
    \label{fig:power2t}
\end{figure}

All findings with respect to power and thrust are summarized in Table~\ref{tab:2T_turb}. Notice that the energy increase at 5$D$ as predicted in Table~\ref{tab:1T_turb} corresponds very well with the actual power increase of a turbine at 5$D$. As a result, the CCW helix approach with a 4$^{\circ}$ pitch amplitude increases the power production of this 2-turbine wind farm with 7.5\%. This is considerably higher than the 4.6\% gain obtained with DIC. The overall energy production of all strategies is shown in Figure~\ref{fig:power2t}.

\begin{table}[b!]
    \centering
    \setlength{\tabcolsep}{4pt} 
    \caption{Turbulent inflow, two-turbine results. All results are relative with respect to the baseline case.}
    \begin{tabular}{l|c|c|c|c|c|c|c|c}
         & \rotatebox[origin=c]{90}{\textbf{Static 1$^\circ$}} & \rotatebox[origin=c]{90}{\textbf{Static 2$^\circ$}} & \rotatebox[origin=c]{90}{\textbf{DIC 2.5$^\circ$}} &  \rotatebox[origin=c]{90}{\textbf{CCW Helix 2.5$^\circ$}} & \rotatebox[origin=c]{90}{\textbf{CW Helix 2.5$^\circ$}} & \rotatebox[origin=c]{90}{\textbf{DIC 4$^\circ$}} & \rotatebox[origin=c]{90}{\textbf{CCW Helix 4$^\circ$}} & \rotatebox[origin=c]{90}{\textbf{CW Helix 4$^\circ$}}  \\ \hline
        \textbf{Power T1} & \color{red}{-1.0\%} & \color{red}{-3.1\%} & \color{red}{-1.1\%} & \color{red}{-1.1\%} & \color{red}{-1.0\%} & \color{red}{-2.8\%}  & \color{red}{-2.8\%} & \color{red}{-2.6\%} \\
        \textbf{Power T2} & +1.6\% & +5.3\% & +14.6\% & +17.2\% & +6.3\% & +27.3\% & +39.5\% & +18.0\%  \\
        \textbf{Total power production} & \color{red}{-0.3\%} & \color{red}{-1.0\%} & +2.8\% & +3.4\% & +0.8\% & +4.6\%  & +7.5\% & +2.5\% \\
        \textbf{Variance of power T1} & -2.2\% & -5.8\% & \color{red}{+79.6\%} & -3.4\% & -1.4\% & \color{red}{+194.0\%} & -7.9\% & -5.1\% \\
        \textbf{Variance of power T2} & -11.0\% & -17.6\% & \color{red}{+280.8\%} & \color{red}{+143.0\%} & \color{red}{+82.2\%} & \color{red}{+583.6\%}  & \color{red}{+239.4\%} & \color{red}{+187.2\%} \\
        \textbf{Variance of thrust T1} & -11.3\% & -22.1\% & \color{red}{+580.7\%} & -1.5\% & \color{red}{+1.1\%} & \color{red}{+1416.7\%}  & -3.9\% & \color{red}{+0.4\%} \\
        \textbf{Variance of thrust T2} & -13.0\% & -25.9\% & \color{red}{+165.1\%} & \color{red}{+71.6\%} & \color{red}{+45.5\%}  & \color{red}{+340.9\%} & \color{red}{+123.1\%} & \color{red}{+99.9\%} 
    \end{tabular}
    \label{tab:2T_turb}
\end{table}

Apart from the power production, it is also interesting to investigate the variations of power and thrust. With both helix approaches, the power and thrust variations of the excited turbine are, in general, slightly reduced. Due to the increased wake velocity and turbulence, the downstream turbines experience a significantly higher power and thrust variations than in the baseline case. However, compared to DIC, these variations are much lower. As a result, the fatigue loads that might lead to structural damage of the wind turbine are expected to be substantially lower than with DIC.


A final note should be made with respect to the performance of the helix approach: as the research presented in this paper serves mainly as a proof of concept, the optimal settings for the helix approach are as of yet unknown. In this study, it was assumed that the optimal excitation frequency is identical to the optimal DIC frequency. As such, the 7.5\% power gain found here can be considered conservative, as a different dynamic input signal might lead to better performance.

\section{Conclusions}\label{sec:conclusions}

In this paper, a novel wind farm control strategy is proposed. The strategy involves using Individual Pitch Control (IPC) to dynamically vary the direction of the thrust force exerted on the flow by a wind turbine, leading to a helical wake that increases mixing. As a result, downstream turbines will experience higher wind speeds and subsequently have a higher power production. Due to the helical shape of the wake, this approach is named the \textit{helix approach}. A proof of concept is given for this novel dynamic wind farm control strategy.

The strategy is tested using high-fidelity LES simulations, proving that the helix approach is effective at increasing wake recovery: the energy in the wake can be increased by up to 47\%. Furthermore, it is observed that a helix rotating in counterclockwise direction results in better wake recovery than a helix rotating in clockwise direction. Simulations with a second turbine in the wake of the controlled turbine, located 5 rotor diameters downstream, show that the energy capture can be increased with up to 7.5\% for this two-turbine wind farm. As the optimal control settings for the helix approach have not yet been evaluated, this gain should be seen as an indication of its potential, not as an upper limit.

The helix approach is compared with different existing control strategies. The current simulations show that it is a more effective method to increase the energy capture of a wind farm than both static derating and dynamic induction control. Compared to the latter, the helix approach results in power and thrust variations that are over a factor 2 lower. Furthermore, unlike yaw-based wake redirection, the operational strategy used in the helix approach does not deviate from the operating range for which the turbine was designed. This should allow for a much quicker adaptation of the technology by the industry, perhaps delivering the first wind farm control methodology that can reliably increase the power production in existing wind farms without the need for slow certification protocols and fundamental turbine redesign.

This paper should be considered as a proof of concept. As the helix approach, or dynamic IPC in general, is a completely novel concept, this paper only shows that it \textit{can} be an effective wind farm control strategy. To determine its full potential, further exploration is necessary. Future research possibilities include, but are limited to, studying the difference between the clockwise and counterclockwise helix, finding the optimal blade excitation signals, investigating the damage equivalent load effects on both the excited and downstream turbine, applying closed-loop control on the yaw and tilt moments, increasing the farm size to study the effect on turbines further downstream, executing scaled wind tunnel experiments and full scale tests on an actual wind turbine or wind farm.

\printendnotes

\bibliography{bibliography}

\begin{thebibliography}{38}
\providecommand{\natexlab}[1]{#1}
\providecommand{\url}[1]{\texttt{#1}}
\providecommand{\urlprefix}{}

\bibitem[{S.~Lissaman(1979)S. Lissaman, PB}]{windfarm1979}
S~Lissaman P.
\newblock Energy effectiveness of arbitrary arrays of wind turbines.
\newblock Journal of Energy 1979;3(6):323--328.

\bibitem[{Jensen(1983)Jensen, Niels Otto}]{jensen}
Jensen NO.
\newblock A note on wind generator interaction; 1983.

\bibitem[{Katic et~al.(1987)Katic, I and H{\o}jstrup, J{\o}rgen and Jensen,
  Niels Otto}]{katic1987}
Katic I, H{\o}jstrup J, Jensen NO.
\newblock A simple model for cluster efficiency.
\newblock In: European wind energy association conference and exhibition; 1987.
  .

\bibitem[{Bastankhah and Port{\'e}-Agel(2016)Bastankhah, Majid and
  Port{\'e}-Agel, Fernando}]{bastankhah}
Bastankhah M, Port{\'e}-Agel F.
\newblock Experimental and theoretical study of wind turbine wakes in yawed
  conditions.
\newblock Journal of Fluid Mechanics 2016;806:506--541.

\bibitem[{Annoni et~al.(2018)Annoni, Jennifer and Fleming, Paul and Scholbrock,
  Andrew and Roadman, Jason and Dana, Scott and Adcock, Christiane and
  Porte-Agel, Fernando and Raach, Steffen and Haizmann, Florian and Schlipf,
  David}]{annonifloris}
Annoni J, Fleming P, Scholbrock A, Roadman J, Dana S, Adcock C, et~al.
\newblock Analysis of control-oriented wake modeling tools using lidar field
  results.
\newblock Wind Energy Science 2018;3(2):819--831.

\bibitem[{Boersma et~al.(2017)Boersma, Sjoerd and Doekemeijer, BM and Gebraad,
  Pieter MO and Fleming, Paul A and Annoni, Jennifer and Scholbrock, Andrew K
  and Frederik, J A and van Wingerden, Jan-Willem}]{boersma2017}
Boersma S, Doekemeijer B, Gebraad PM, Fleming PA, Annoni J, Scholbrock AK,
  et~al.
\newblock A tutorial on control-oriented modeling and control of wind farms.
\newblock In: 2017 American Control Conference (ACC) IEEE; 2017. p. 1--18.

\bibitem[{Marden et~al.(2013)Marden, Jason R and Ruben, Shalom D and Pao, Lucy
  Y}]{marden}
Marden JR, Ruben SD, Pao LY.
\newblock A model-free approach to wind farm control using game theoretic
  methods.
\newblock IEEE Transactions on Control Systems Technology
  2013;21(4):1207--1214.

\bibitem[{Ciri et~al.(2017)Ciri, Umberto and Rotea, Mario A and Leonardi,
  Stefano}]{rotea}
Ciri U, Rotea MA, Leonardi S.
\newblock Model-free control of wind farms: A comparative study between
  individual and coordinated extremum seeking.
\newblock Renewable energy 2017;113:1033--1045.

\bibitem[{Vali et~al.(2016)Vali, Mehdi and van Wingerden, Jan-Willem and
  Boersma, Sjoerd and Petrovi{\'c}, Vlaho and K{\"u}hn, M}]{vali}
Vali M, van Wingerden JW, Boersma S, Petrovi{\'c} V, K{\"u}hn M.
\newblock A predictive control framework for optimal energy extraction of wind
  farms.
\newblock In: Journal of Physics: Conference Series, vol. 753 IOP Publishing;
  2016. p. 052013.

\bibitem[{Annoni et~al.(2016)Annoni, Jennifer and Gebraad, Pieter MO and
  Scholbrock, Andrew K and Fleming, Paul A and Wingerden, Jan-Willem
  van}]{annoni}
Annoni J, Gebraad PM, Scholbrock AK, Fleming PA, Wingerden JWv.
\newblock Analysis of axial-induction-based wind plant control using an
  engineering and a high-order wind plant model.
\newblock Wind Energy 2016;19(6):1135--1150.

\bibitem[{Campagnolo et~al.(2016)F. Campagnolo and V. Petrovi\'c and C. L.
  Bottasso and A. Croce}]{Campagnolo2016a}
Campagnolo F, Petrovi\'c V, Bottasso CL, Croce A.
\newblock Wind tunnel testing of wake control strategies.
\newblock In: 2016 American Control Conference (ACC); 2016. p. 513--518.

\bibitem[{van~der Hoek et~al.(2019)van der Hoek, Daan and Kanev, Stoyan and
  Allin, Julian and Bieniek, David and Mittelmeier, Niko}]{vdhoek2019}
van~der Hoek D, Kanev S, Allin J, Bieniek D, Mittelmeier N.
\newblock Effects of axial induction control on wind farm energy production-A
  field test.
\newblock Renewable Energy 2019;140:994--1003.

\bibitem[{Jim{\'e}nez et~al.(2010)Jim{\'e}nez, {\'A}ngel and Crespo, Antonio
  and Migoya, Emilio}]{jimenez2010}
Jim{\'e}nez {\'A}, Crespo A, Migoya E.
\newblock Application of a LES technique to characterize the wake deflection of
  a wind turbine in yaw.
\newblock Wind energy 2010;13(6):559--572.

\bibitem[{Raach et~al.(2016)Raach, Steffen and Schlipf, David and Cheng, Po
  Wen}]{raach2016}
Raach S, Schlipf D, Cheng PW.
\newblock Lidar-based wake tracking for closed-loop wind farm control.
\newblock In: Journal of Physics: Conference Series, vol. 753 IOP Publishing;
  2016. p. 052009.

\bibitem[{Campagnolo et~al.(2016)Filippo Campagnolo and Vlaho Petrovi\'c and
  Johannes Schreiber and Emmanouil M. Nanos and Alessandro Croce and Carlo L.
  Bottasso}]{Campagnolo2016_WFC}
Campagnolo F, Petrovi\'c V, Schreiber J, Nanos EM, Croce A, Bottasso CL.
\newblock Wind tunnel testing of a closed-loop wake deflection controller for
  wind farm power maximization.
\newblock Journal of Physics: Conference Series 2016;753(3):7.

\bibitem[{Fleming et~al.(2017)Fleming, Paul and Annoni, Jennifer and Shah,
  Jigar J and Wang, Linpeng and Ananthan, Shreyas and Zhang, Zhijun and
  Hutchings, Kyle and Wang, Peng and Chen, Weiguo and Chen, Lin}]{fleming2017}
Fleming P, Annoni J, Shah JJ, Wang L, Ananthan S, Zhang Z, et~al.
\newblock Field test of wake steering at an offshore wind farm.
\newblock Wind Energy Science 2017;2(1):229--239.

\bibitem[{Howland et~al.(2019)Howland, Michael F and Lele, Sanjiva K and
  Dabiri, John O}]{howland}
Howland MF, Lele SK, Dabiri JO.
\newblock Wind farm power optimization through wake steering.
\newblock Proceedings of the National Academy of Sciences
  2019;116(29):14495--14500.

\bibitem[{Westergaard(2013)Westergaard, C H}]{dicpatent}
Westergaard CH, A method for improving large array wind park power performance
  through active wake manipulation reducing shadow effects; 2013.

\bibitem[{Munters and Meyers(2016)Munters, Wim and Meyers,
  Johan}]{Munters:2016}
Munters W, Meyers J.
\newblock Effect of wind turbine response time on optimal dynamic induction
  control of wind farms.
\newblock In: Journal of Physics: Conference Series, vol. 753 IOP Publishing;
  2016. p. 052007.

\bibitem[{Munters and Meyers(2017)Munters, Wim and Meyers,
  Johan}]{Munters:2017}
Munters W, Meyers J.
\newblock An optimal control framework for dynamic induction control of wind
  farms and their interaction with the atmospheric boundary layer.
\newblock Phil Trans R Soc A 2017;375(2091):20160100.

\bibitem[{Munters and Meyers(2018)Munters, Wim and Meyers,
  Johan}]{Munters:2018}
Munters W, Meyers J.
\newblock Towards practical dynamic induction control of wind farms: analysis
  of optimally controlled wind-farm boundary layers and sinusoidal induction
  control of first-row turbines.
\newblock Wind Energy Science 2018;3(1):409--425.

\bibitem[{Frederik et~al.(2019)Frederik, Joeri A and Campagnolo, Filippo and
  Cacciola, Stefano and Weber, Robin and Croce, Alessandro and Bottasso, Carlo
  and van Wingerden, Jan-Willem}]{frederik2019}
Frederik JA, Campagnolo F, Cacciola S, Weber R, Croce A, Bottasso C, et~al.
\newblock Periodic dynamic induction control of wind farms: proving the
  potential in simulations and wind tunnel experiments.
\newblock Wind Energy Science 2019;under review.

\bibitem[{Kimura et~al.(2019)Kimura, Keita and Tanabe, Yasutada and Matsuo,
  Yuichi and Iida, Makoto}]{dicyaw}
Kimura K, Tanabe Y, Matsuo Y, Iida M.
\newblock Forced wake meandering for rapid recovery of velocity deficits in a
  wind turbine wake.
\newblock In: AIAA Scitech 2019 Forum; 2019. p. 2083.

\bibitem[{Bossanyi(2003)Bossanyi, Ervin A}]{bossanyi2003individual}
Bossanyi EA.
\newblock Individual blade pitch control for load reduction.
\newblock Wind Energy: An International Journal for Progress and Applications
  in Wind Power Conversion Technology 2003;6(2):119--128.

\bibitem[{Bossanyi(2005)Bossanyi, Ervin A}]{bossanyi2005further}
Bossanyi EA.
\newblock Further load reductions with individual pitch control.
\newblock Wind Energy: An International Journal for Progress and Applications
  in Wind Power Conversion Technology 2005;8(4):481--485.

\bibitem[{Mulders et~al.(2019)Mulders, Sebastiaan Paul and Pamososuryo,
  Atindriyo Kusumo and Disario, Gianmarco Emilio and {van Wingerden}, Jan
  Willem}]{mulders2019analysis}
Mulders SP, Pamososuryo AK, Disario GE, {van Wingerden} JW.
\newblock Analysis and optimal individual pitch control decoupling by inclusion
  of an azimuth offset in the multiblade coordinate transformation.
\newblock Wind Energy 2019;22(3):341--359.

\bibitem[{Navalkar et~al.(2014)Navalkar, Sachin T and van Wingerden, Jan-Willem
  and van Solingen, Edwin and Oomen, Tom and Pasterkamp, Edwin and Van Kuik,
  GAM}]{navalkar2014subspace}
Navalkar ST, van Wingerden JW, van Solingen E, Oomen T, Pasterkamp E, Van~Kuik
  G.
\newblock Subspace predictive repetitive control to mitigate periodic loads on
  large scale wind turbines.
\newblock Mechatronics 2014;24(8):916--925.

\bibitem[{Frederik et~al.(2018{\natexlab{a}})Frederik, Joeri A and Kr{\"o}ger,
  Lars and G{\"u}lker, Gerd and van Wingerden, Jan-Willem}]{frederik2018cep}
Frederik JA, Kr{\"o}ger L, G{\"u}lker G, van Wingerden JW.
\newblock Data-driven repetitive control: Wind tunnel experiments under
  turbulent conditions.
\newblock Control Engineering Practice 2018;80:105--115.

\bibitem[{Frederik et~al.(2018{\natexlab{b}})Frederik, Joeri A and Kr{\"o}ger,
  Lars and Peinke, Joachim and H{\"o}lling, Michael and van Wingerden,
  Jan-Willem}]{frederik2018torque}
Frederik JA, Kr{\"o}ger L, Peinke J, H{\"o}lling M, van Wingerden JW.
\newblock Validating subspace predictive repetitive control under turbulent
  wind conditions with wind tunnel experiment.
\newblock In: Journal of Physics: Conference Series, vol. 1037 IOP Publishing;
  2018. p. 032008.

\bibitem[{Fleming et~al.(2014)Fleming, Paul A and Gebraad, Pieter MO and Lee,
  Sang and van Wingerden, Jan-Willem and Johnson, Kathryn and Churchfield, Matt
  and Michalakes, John and Spalart, Philippe and Moriarty,
  Patrick}]{ipcwakesteering}
Fleming PA, Gebraad PM, Lee S, van Wingerden JW, Johnson K, Churchfield M,
  et~al.
\newblock Evaluating techniques for redirecting turbine wakes using SOWFA.
\newblock Renewable Energy 2014;70:211--218.

\bibitem[{Fleming et~al.(2015)Fleming, Paul and Gebraad, Pieter MO and Lee,
  Sang and van Wingerden, Jan-Willem and Johnson, Kathryn and Churchfield, Matt
  and Michalakes, John and Spalart, Philippe and Moriarty,
  Patrick}]{fleming2015ipc}
Fleming P, Gebraad PM, Lee S, van Wingerden JW, Johnson K, Churchfield M,
  et~al.
\newblock Simulation comparison of wake mitigation control strategies for a
  two-turbine case.
\newblock Wind Energy 2015;18(12):2135--2143.

\bibitem[{van Wingerden et~al.(2019)van Wingerden, J W and Frederik, J.A. and
  Doekemeijer B M}]{ipcpatent}
van Wingerden JW, Frederik JA, M DB, Enhanced wind turbine wake mixing; 2019.

\bibitem[{Churchfield and Lee(2012)Churchfield, M and Lee, S}]{SOWFA_General}
Churchfield M, Lee S.
\newblock NWTC design codes-SOWFA.
\newblock 15013 Denver West Parkway, Golden, CO.
  http://wind.nrel.gov/designcodes/simulators/SOWFA: National Renewable Energy
  Laboratory (NREL); 2012.

\bibitem[{Churchfield et~al.(2012)Matthew J. Churchfield and Sang Lee and John
  Michalakes and Patrick J. Moriarty}]{Churchfield2012}
Churchfield MJ, Lee S, Michalakes J, Moriarty PJ.
\newblock A numerical study of the effects of atmospheric and wake turbulence
  on wind turbine dynamics.
\newblock Journal of Turbulence 2012;13:N14.

\bibitem[{S{\o}rensen and Shen(2002)S{\o}rensen, Jens N{\o}rk{\ae}r and Shen,
  Wen Zhong}]{Sorensen2012}
S{\o}rensen JN, Shen WZ.
\newblock {Numerical Modeling of Wind Turbine Wakes }.
\newblock Journal of Fluids Engineering 2002 05;124(2):393--399.

\bibitem[{Bak et~al.(2013)Bak, C and Zhale, F and Bitsche, R and Kim, T and
  Yde, A and Henriksen, L C and Hansen, M H and Blasques, J P A A and Gaunaa, M
  abd Natarajan, A}]{DTU10MW}
Bak C, Zhale F, Bitsche R, Kim T, Yde A, Henriksen LC, et~al.
\newblock The DTU 10-MW Reference Wind Turbine.
\newblock Technical University of Denmark (DTU), Department of Wind Energy.
  http://dtu-10mw-rwt.vindenergi.dtu.dk/; 2013.

\bibitem[{Nilsson et~al.(2015)Nilsson, Karl and Ivanell, Stefan and Hansen,
  Kurt S and Mikkelsen, Robert and S{\o}rensen, Jens N and Breton,
  Simon-Philippe and Henningson, Dan}]{nilsson}
Nilsson K, Ivanell S, Hansen KS, Mikkelsen R, S{\o}rensen JN, Breton SP, et~al.
\newblock Large-eddy simulations of the Lillgrund wind farm.
\newblock Wind Energy 2015;18(3):449--467.

\bibitem[{Bir(2008)Bir, Gunjit}]{mbc}
Bir G.
\newblock Multi-blade coordinate transformation and its application to wind
  turbine analysis.
\newblock In: 46th AIAA aerospace sciences meeting and exhibit; 2008. p. 1300.

\end{thebibliography}



\end{document}